\documentclass[12pt]{article}

\usepackage{newtxtext,newtxmath}
\usepackage{graphicx}
\usepackage{multirow}
\usepackage[letterpaper,margin=1in]{geometry}

\renewenvironment{abstract}
	{\quotation}
	{\endquotation}

\date{}

\makeatletter
\renewcommand{\fnum@figure}{\textbf{Figure \thefigure}}
\renewcommand{\fnum@table}{\textbf{Table \thetable}}
\makeatother

\usepackage{scicite}

\usepackage{url}

\def\scititle{Ultra-high-energy $\gamma$-ray imprints from PeV particles accelerated by supernova remnants
}
\title{\bfseries \boldmath \scititle}

\author{\noindent Authors and affiliations appear at the end of the paper.}
\begin{document} 

% Insert the title and author list
\maketitle

% Abstract, in bold
% There are strict length limits, and not all formats have abstracts.
% Consult the journal instructions to authors for details.
% Do not cite any references in the abstract.
\begin{abstract} \bfseries \boldmath
% Start with one or two sentences of background
%This is a simple template to prepare papers in \LaTeX\ for the \textit{Science}-family journals.
%Abstracts start with one or two sentences of background, which should be
%comprehensible to any scientist.
% Then summarise the results of your observations, experiments, simulations etc.
%The following text should outline the main results of the research.
%Simple mathematical expressions can be included e.g. $a^2+b^2=c^2$.
% End with a statement of your main conclusions
%The final sentence of the abstract should state the main conclusions and implications.
The quest for the origin of cosmic ray (CRs) is a fundamental issue in 
astrophysics\cite{1976RvMP...48..161G,2013A&ARv..21...70B}. 
Shocks of supernova remnants (SNRs) have been considered as the dominant contributors 
to Galactic CRs below the spectral knee near $\sim 3$ petaelectronvolt (PeV). 
Whether SNRs are efficient accelerators of particles beyond PeV energies has long been debated\cite{1983A&A...125..249L, 2013MNRAS.431..415B}. 
Here we report observations of very-high-energy $\gamma$-ray emission up to hundreds 
of TeV from two middle age shell-type SNRs, G150.3$+$4.5 and $\gamma$-Cygni, with the 
Large High Altitude Air Shower Observatory (LHAASO). Two (or three) distinct 
morphological/spectral components with convex spectral shapes are observed in both 
sources, with the low-energy one being more extended than the high-energy one. 
%Although it is possible that these high-energy components may be driven by powerful pulsars, 
The likely association of the high-energy component with molecular clouds at similar 
distances, and the weakness/absence of pulsar wind nebulae (PWNe) inside these
SNRs\cite{2015ApJ...799...76H, 2024A&A...689A.257L} clearly indicate for the first time that the highest energy emission is produced 
by collision of hadronic CRs up to PeV energies with the clouds. 
These results are compatible with the classic model prediction that PeV 
particles accelerated near the end of the free expansion phase of SNR evolution 
can illuminate nearby molecular clouds (MCs) to produce strong $\gamma$-ray
emission\cite{1994A&A...285..645A,1996A&A...309..917A}.
\end{abstract}

% The first paragraph of any Science paper does NOT have a heading
% Nor is it indented
\noindent
It is generally accepted that SNRs shocks provide sufficient energy to sustain the
observed GeV CR fluxes\cite{1934PhRv...46...76B}. Evidence of a neutral pion decay signature 
in several SNRs accompanied with MCs suggests that SNRs are hadronic particle 
accelerators\cite{2010ApJ...710L.151T,2013Sci...339..807A}, supporting the scenario of 
SNR origin of Galactic CRs. However, there is no direct evidence that SNRs act as PeV 
particle accelerators (hereafter PeVatrons). The unprecedented sensitivity to ultra-high-energy
 $\gamma$-ray observations of the Large High Altitude Air Shower Observatory
(LHAASO)\cite{2022ChPhC..46c0001M} offers a unique opportunity to address the question 
of the maximum acceleration energy of SNRs\cite{2024SciBu..69.2833C} and their contributions 
to CRs up to the knee region where a prominent break is seen\cite{2024PhRvL.132m1002C}.

In an earlier study, ultra-high-energy $\gamma$-ray emission was reported from the compact SNR W51 with a powerful PWN inside \cite{2024SciBu..69.2833C}. 
%\cite{2011ApJ...728L..28E, 2021Univ....7..324C} and there are indications that cosmic rays trapped in SNRs have a very soft spectrum above $\sim 0.1$PeV \cite{2021ApJ...910...78Z}. The unprecedented sensitivity of the  Large High Altitude Air Shower Observatory (LHAASO) in the ultra-high-energy range has made it possible to extend the $\gamma$-ray spectral measurement beyond the PeV energy \cite{2021Natur.594...33C, 2021Sci...373..425L, 2022ChPhC..46c0001M}. 
Here we report observations of two SNRs, G150.3$+$4.5 and $\gamma$-Cygni, with LHAASO.
The SNR G150.3$+$4.5 was first identified by radio observations, showing a radius of about 
$1.3^\circ$\cite{2014A&A...566A..76G}. Recent CO observations show that the distance from the 
SNR is about 740 pc and that the age is estimated to be $(1-18)\times10^4$ yr\cite{2024A&A...686A.305F}. 
Using the statistical diameter-age relation\cite{2023ApJS..265...53R}, an age of 26 kyr 
can be derived. The overall GeV $\gamma$-ray spectrum is hard, with a photon index of 
$-1.6$ at 9 GeV with a significant spectral curvature toward 
high energies\cite{2020A&A...643A..28D}. Detailed analysis of the Fermi data 
showed that the $\gamma$-ray spectra have spatial variations, suggesting a hybrid 
radiation process\cite{2024A&A...689A.257L}. In the first catalog of LHAASO, an extended 
source associated with the radio shell around TeV energies and a more compact component 
at higher energies were detected\cite{2024ApJS..271...25C}.

SNR $\gamma$-Cygni (G78.2$+$2.1 or DR4) has been extensively studied in the radio, X-ray, 
and $\gamma$-ray bands\cite{2013MNRAS.436..968L,2015ApJ...799...76H,2013ApJ...770...93A,
2018ApJ...861..134A,2023A&A...670A...8M}. The age of this SNR is estimated to be about 
7 kyr, and the distance is $\sim 1.7$ kpc\cite{2000AstL...26...77L}. Its radio shell has 
an angular radius of about $0.5^{\circ}$, which corresponds to a scale of $\sim15$ pc. 
%\textbf{Observations show that the shock speed is $\sim 1000$ km s$^{-1}$~\cite{2023A&A...670A...8M}, 
%suggesting that it may accelerate CR particles efficiently.} 
Strong $\gamma$-ray emission 
has been detected from the pulsar PSR J2021+4026, which is a bit off to the east of the 
SNR center and might be associated with the SNR\cite{2016ApJ...825...18N}. However, the 
PWN of PSR J2021+4026 is very weak in radio and X-ray bands\cite{2015ApJ...799...76H}, 
which %,\textbf{delete this: in combination with the young age of the SNR and strong magnetic field inferred from X-ray observations,} 
implies weak TeV emission around the pulsar. This source was 
reported in the first LHAASO catalog\cite{2024ApJS..271...25C}, and was associated with 
VER J2019+407 as detected by VERITAS\cite{2013ApJ...770...93A}.

\section*{LHAASO observations}

\begin{figure}[!htb]
\centering
\includegraphics[width=0.32\textwidth]{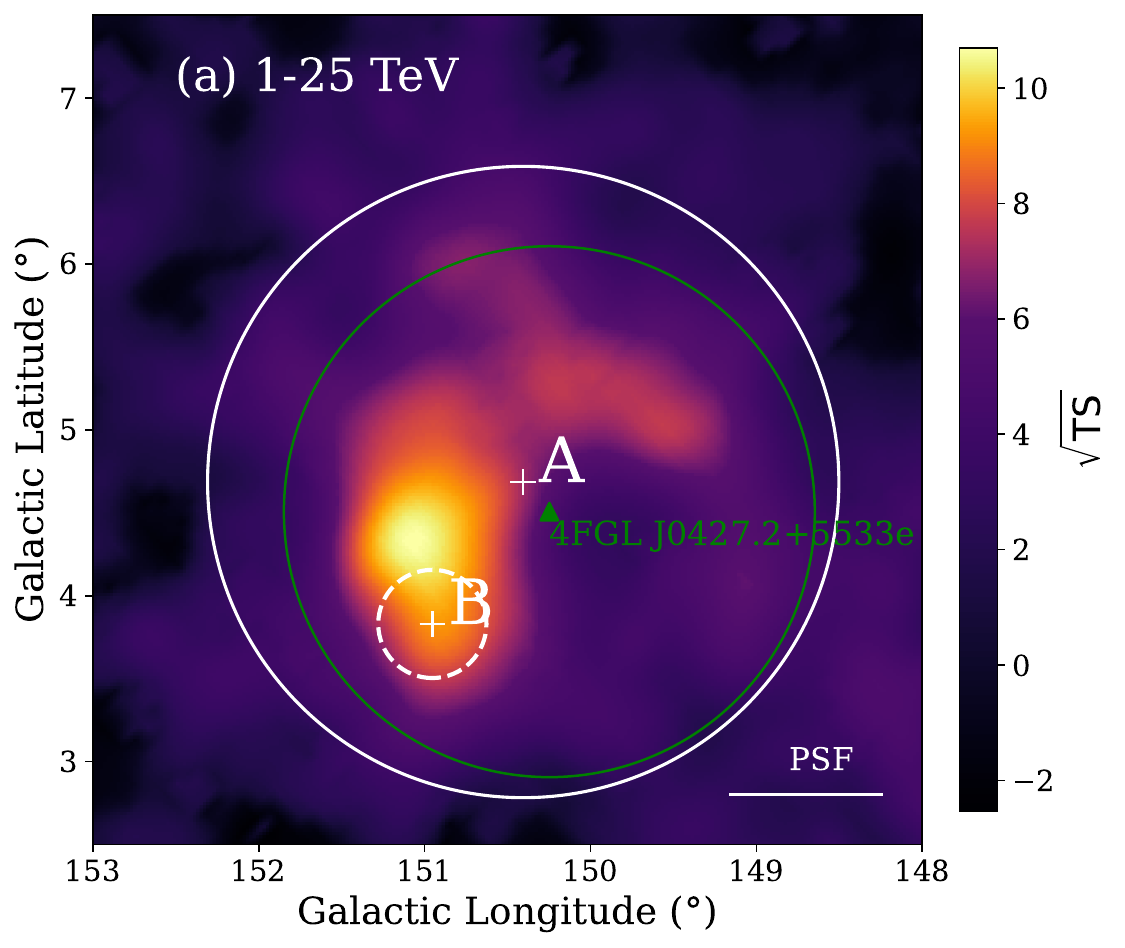}
\includegraphics[width=0.32\textwidth]{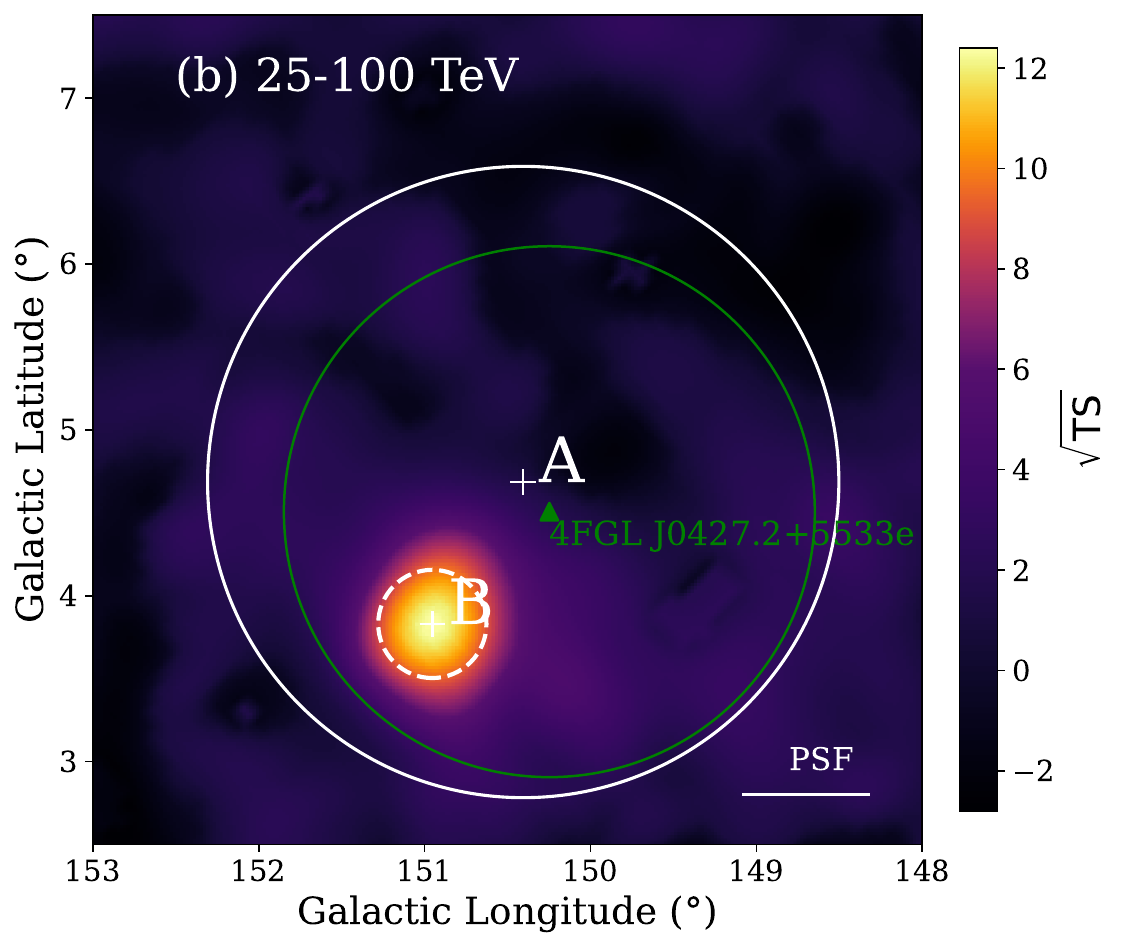}
\includegraphics[width=0.32\textwidth]{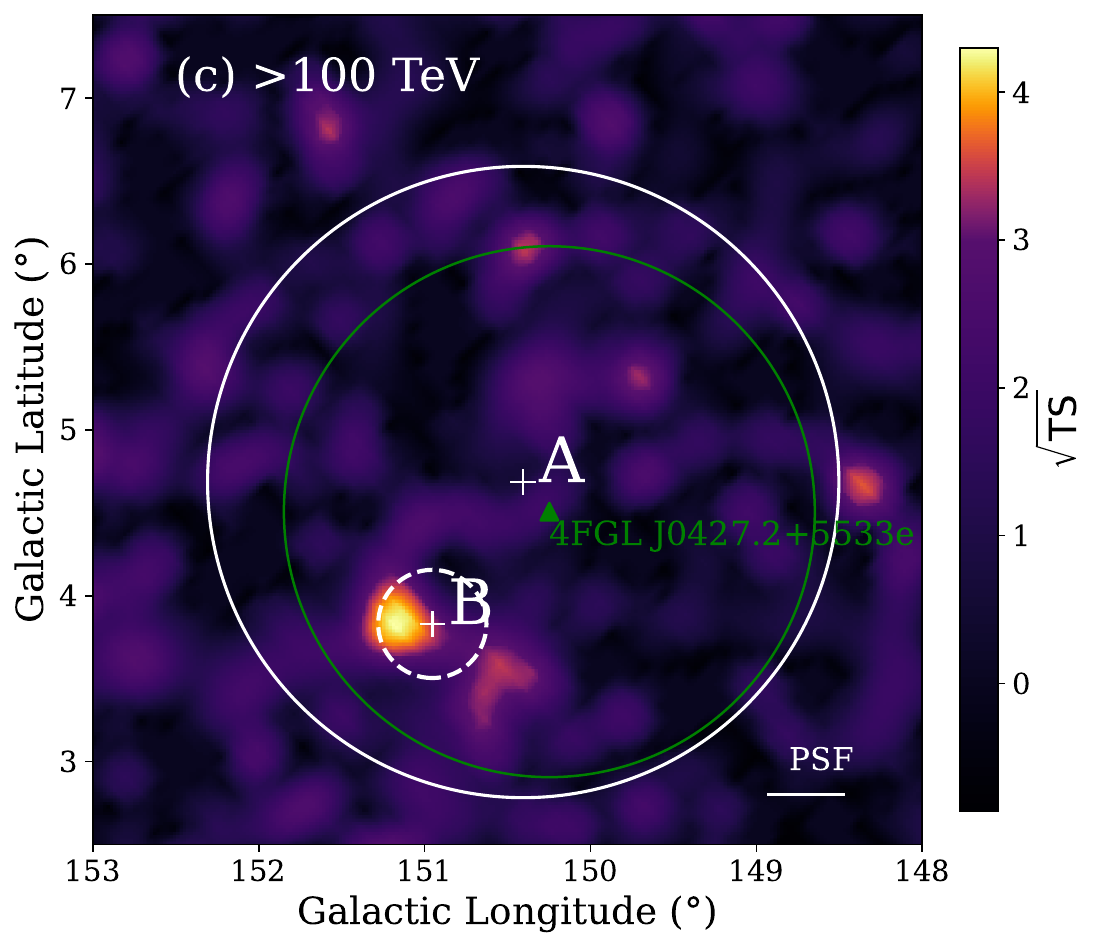}
\includegraphics[width=0.32\textwidth]{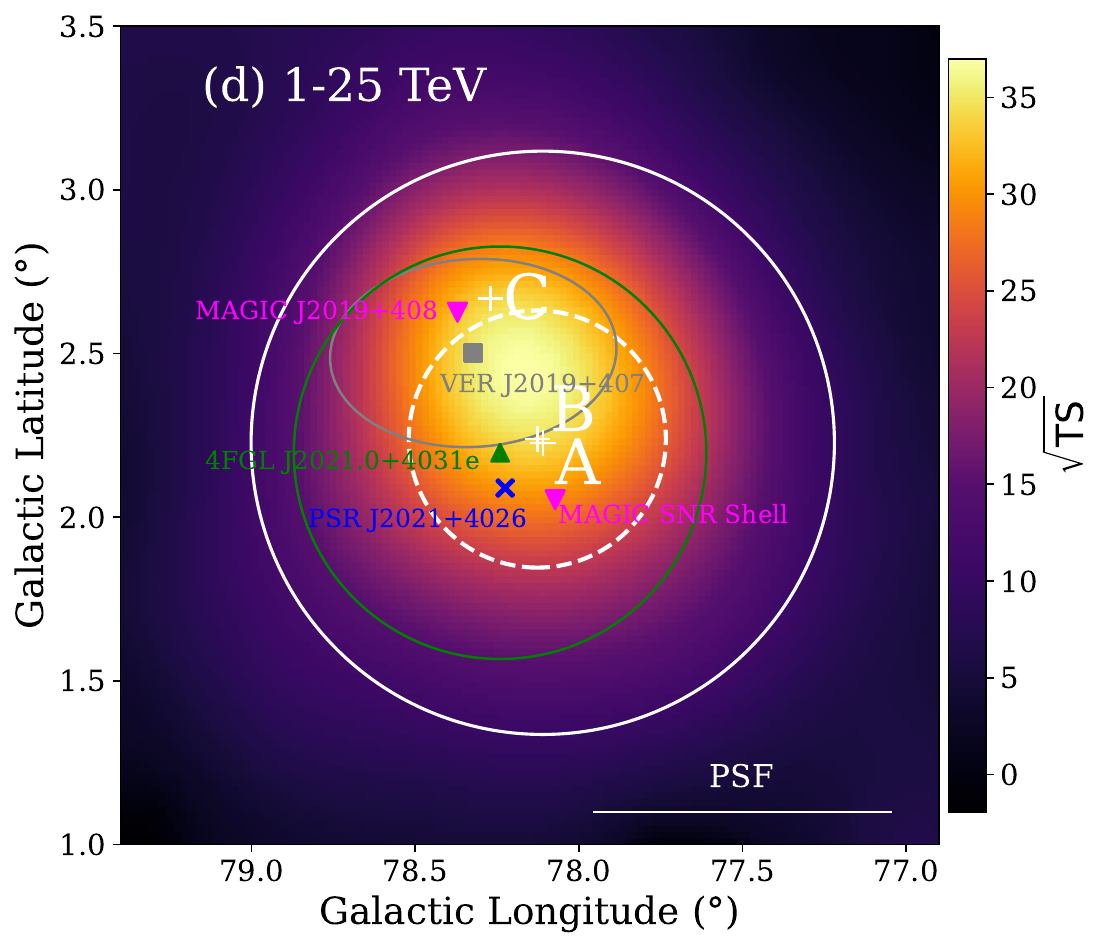}
\includegraphics[width=0.32\textwidth]{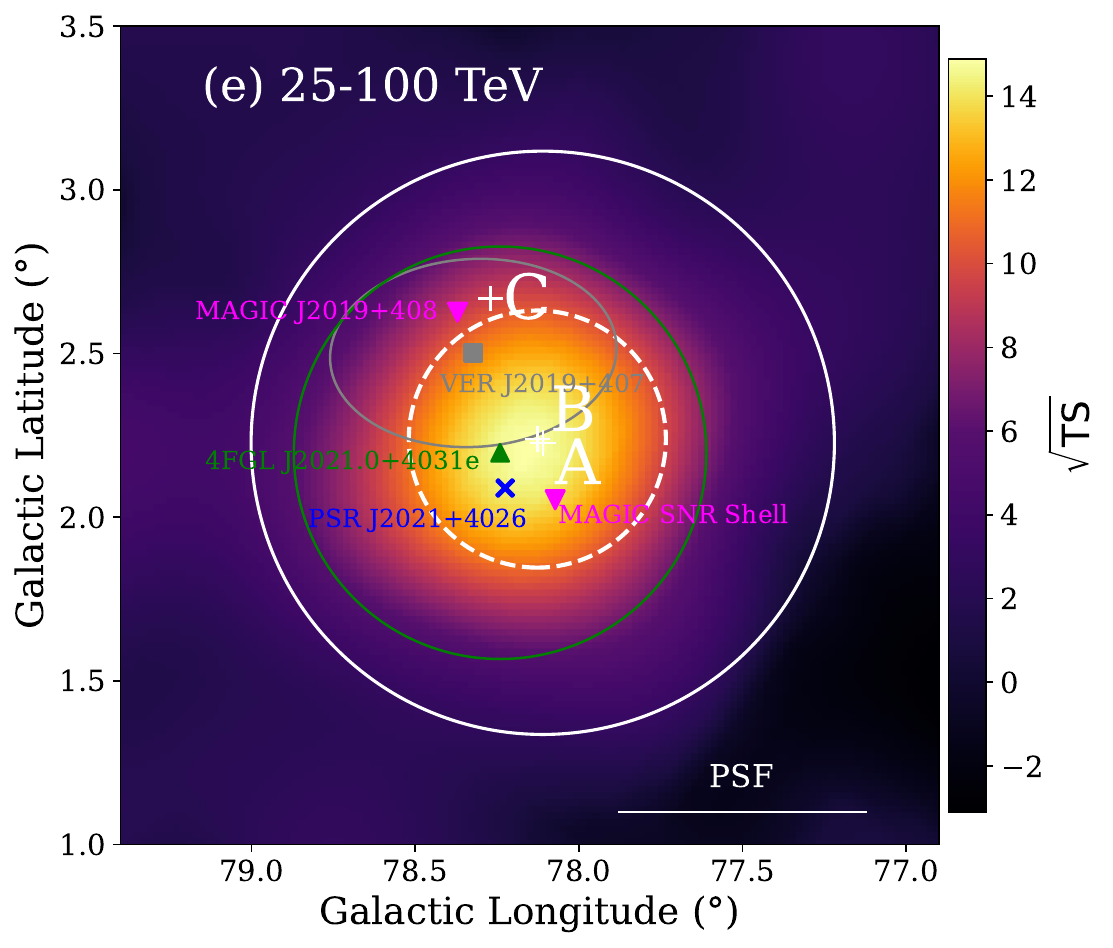}
\includegraphics[width=0.32\textwidth]{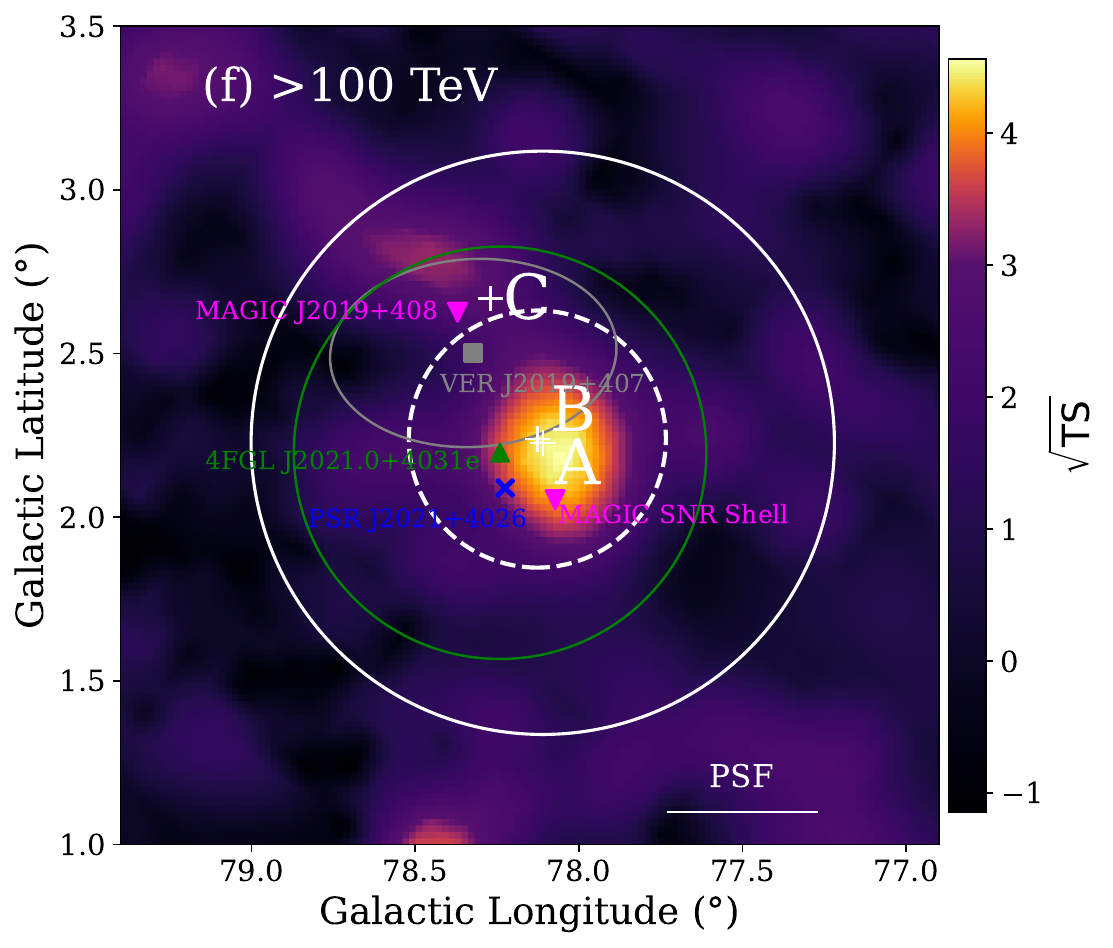}
\caption{{\bf Significance maps around the two SNRs.}
Top panels are for SNR G150.3$+$4.5, and bottom panels are for SNR $\gamma$-Cygni. The panels 
from left to right correspond to three energy bands: $1-25$ TeV, $25-100$ TeV, and $>100$ TeV,
respectively. The white line segment in each panel illustrates the diameter of the PSF ($68\%$ containment).
The best-fit centroid locations and the $68\%$ extensions of LHAASO sources (labeled as A, B,
and C) are indicated by white crosses and circles (solid for A and dashed for B), respectively. 
%Due to the close proximity of the A and B components of $\gamma$-Cygni, they are marked with open diamonds. 
Also marked are locations of sources detected by other experiments (green for Fermi-LAT in the GeV band, gray for VERITAS and magenta for MAGIC in the TeV band, blue for pulsar). 
}
\label{fig:sigmap}
\end{figure}

Using more than three years of data recorded by LHAASO, we study these two SNRs in detail.
The $\gamma$-ray spectrum and morphology are fitted simultaneously using a 3-dimensional 
likelihood fitting method (two for the spatial distribution and one for the spectrum).
We assume a two-dimensional Gaussian morphology and power-law with exponential cutoff 
(PLEcut) spectrum to describe the target source. In total, there are 6 free parameters 
for one source. To characterize the significance of target source detection, a 
likelihood-ratio test is performed, with the test statistic (TS) defined as 
TS~$=2\ln({\mathcal L}_{s+b}/{\mathcal L}_b)$, where ${\mathcal L}_{s+b}$ is the maximum 
likelihood for the signal plus background hypothesis and ${\mathcal L}_b$ is the likelihood 
for the background only hypothesis. Figure~\ref{fig:sigmap} shows the significance maps 
(corresponding to $\sqrt{{\rm TS}}$) of the two SNRs (the top panels are for G150.3$+$4.5 and 
the bottom panels are for $\gamma$-Cygni) in three energy bands of $1-25$, $25-100$ 
and $>100$ TeV. The total detection significance of these two sources is 
$29.8\sigma$ (for G150.3$+$4.5) and $48.9\sigma$ (for $\gamma$-Cygni). Above 100 TeV, we 
still detect $\gamma$-ray emission at $4.3\sigma$ (for G150.3$+$4.5 region) and $4.5\sigma$ 
(for $\gamma$-Cygni region). 

As can be seen in Figure~\ref{fig:sigmap}, the morphology of the $\gamma$-ray emission
changes with energy. This is most evident for SNR G150.3$+$4.5. In the low-energy
band ($1-25$ TeV), the emission of G150.3$+$4.5 is much more extended than in the 
high-energy band ($>25$ TeV). The $1-25$ TeV emission map is in good agreement with the 
radio image of the SNR\cite{2014A&A...566A..76G} and the well-resolved Fermi-LAT $\gamma$ -ray 
map in the GeV band\cite{2020A&A...643A..28D,2024A&A...689A.257L}. The high-energy emission 
concentrates on a southern compact region.
For $\gamma$-Cygni, with increasing energy, the extension gradually decreases, 
and the emission peak shifts toward the SNR center. 
%For $\gamma$-Cygni, the emission is also more extended in the low-energy band ($\sigma=xxx$ in $1-25$ TeV for a two-dimensional Gaussian fitting) than those in the high-energy band ($\sigma=yyy$ in $25-100$ TeV and $zzz$ for $E>100$ TeV). In addition, the emission centroid shifts to the south with increasing energy. 
%besides a point source associated with VER J2019+407 and/or MAGIC J2019+408 (C), the emission is more extended in the low-energy band (A: $0.59^{\circ}$ Gaussian width) than that in the high-energy band (B: $0.26^{\circ}$ Gaussian width).

Therefore, we used multiple sources to fit the data and find that, for G150.3$+$4.5, the 
two-source model produces an improvement of 143 in the TS value compared to the one-source 
model. A second source component is favored with a significance of $11\sigma$ for 6 
more free parameters. The source dominating at low energies ($1-25$ TeV), labeled source 
A, shows an extension of $1.26^{\circ}$ (2D Gaussian width, namely $R_{39}$), and the other (source B) 
dominating at high energies ($>25$ TeV) has a much smaller extension of $0.22^{\circ}$. 
For $\gamma$-Cygni, we find that the emission can be well described by three sources: 
source A with a Gaussian width of $0.59^{\circ}$ is more evident at low energies ($1-25$ TeV), 
the positionally consistent source B with a Gaussian width of $0.26^{\circ}$ dominates at high energies ($>25$ TeV), and 
a point-like source C is positionally consistent with VER J2019+407\cite{2013ApJ...770...93A} 
and/or MAGIC J2019+408\cite{2023A&A...670A...8M}. The TS value increases by $114$ compared to 
the single-source fitting and by $61$ compared to the two-source fitting. The fitting results 
of the source locations and extensions are shown in Figure \ref{fig:sigmap} and are compiled in Table
\ref{tab:morphology}, where GL and GB are the Galactic longitude and latitude, respectively.

\begin{table}[htb]
\small
\centering
\caption{\textbf{Fitting locations and $39\%$ containment radii of multiple source components
of the two SNRs.
%TS values, and additional numbers of free parameters compared with the background hypothesis of one- and two-source assumptions.
}}
\begin{tabular}{c|c|c|c|c|c} \hline \hline
Source & Component &  GL ($^\circ$) & GB ($^\circ$) & $R_{39}$ ($^\circ$) &TS  \\ \hline
\multirow{2}{*}{G150.3+4.5}
  & A & $150.41\pm0.09$ & $4.69\pm0.10$ & $1.26\pm0.07$  &  466\\
  & B & $150.96\pm0.04$ & $3.83\pm0.04$ & $0.22\pm0.04$  & 172 \\ \hline
\multirow{3}{*}{$\gamma$-Cygni}  
& A & $78.11\pm0.07$ & $2.23\pm0.05$ & $0.59\pm0.03$   &632\\  
& B & $78.13\pm0.05$ & $2.24\pm0.03$  & $0.26\pm0.02$  &499\\  
&  C & $78.27\pm0.10$ & $2.66\pm0.05$ & Point Source & 182\\

\hline 
\end{tabular}
\label{tab:morphology}
\end{table}

\iffalse
\begin{table}[htb]
\small
\centering
\caption{\textbf{Fitting locations and $39\%$ containment radii of multiple source components
of the two SNRs.
%TS values, and additional numbers of free parameters compared with the background hypothesis of one- and two-source assumptions.
}}
\begin{tabular}{c|c|c|c|c|c|c|c} \hline \hline
Source & Component &  R.A. ($^\circ$) & Dec. ($^\circ$) & $R_{39}$ ($^\circ$) &TS & Galactic  \\ \hline
\multirow{2}{*}{G150.3+4.5}
  & A & $67.15\pm0.17$ & $55.51\pm0.09$ & $1.26\pm0.07$  & ts & (150,4.5)\\
  & B & $66.73\pm0.07$ & $54.52\pm0.04$ & $0.22\pm0.04$   \\ \hline
\multirow{3}{*}{$\gamma$-Cygni}  
& A & $305.14\pm0.08$ & $40.43\pm0.05$ & $0.59\pm0.03$ \\  
& B & $305.14\pm0.06$ & $40.45\pm0.03$  & $0.26\pm0.02$ \\  
&  C & $304.79\pm0.06$ & $40.81\pm0.11$ & Point Source \\
\hline 
\end{tabular}
\label{tab:morphology}
\end{table}
\fi

\begin{figure}[!htb]
\centering
\includegraphics[width=0.32\textwidth]{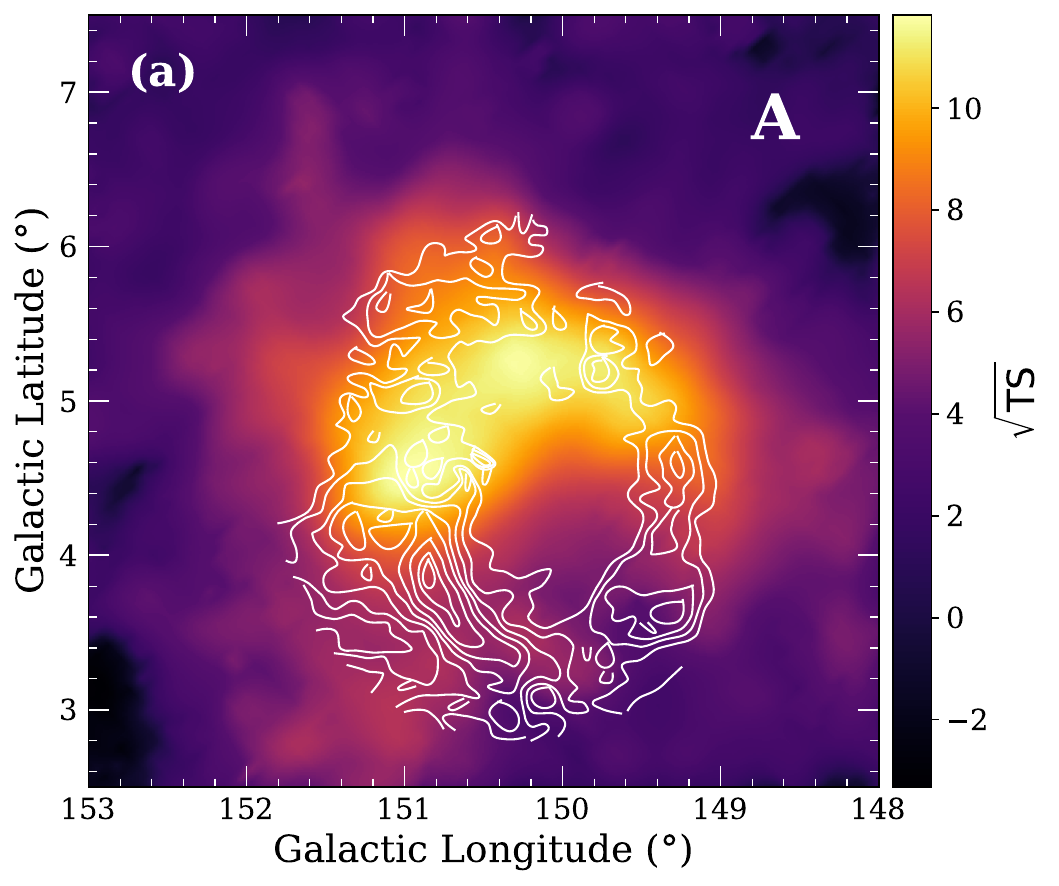}
\includegraphics[width=0.32\textwidth]{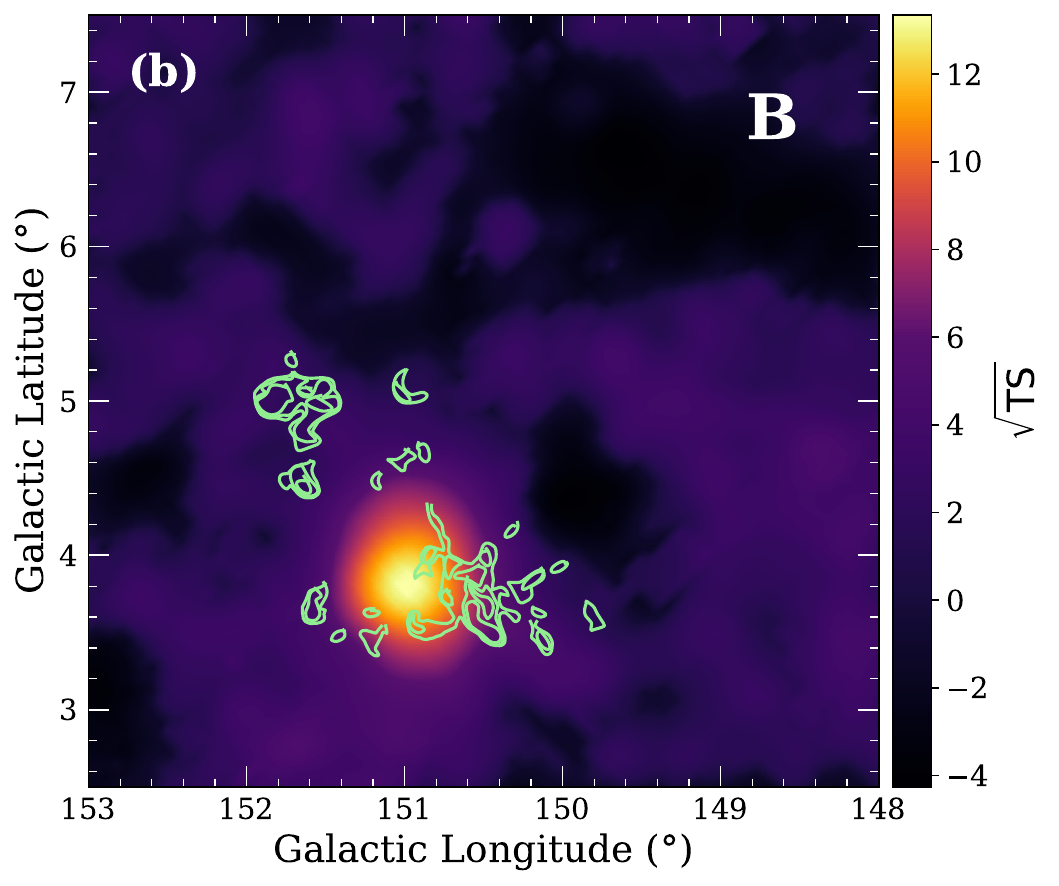}
\includegraphics[width=0.32\textwidth]{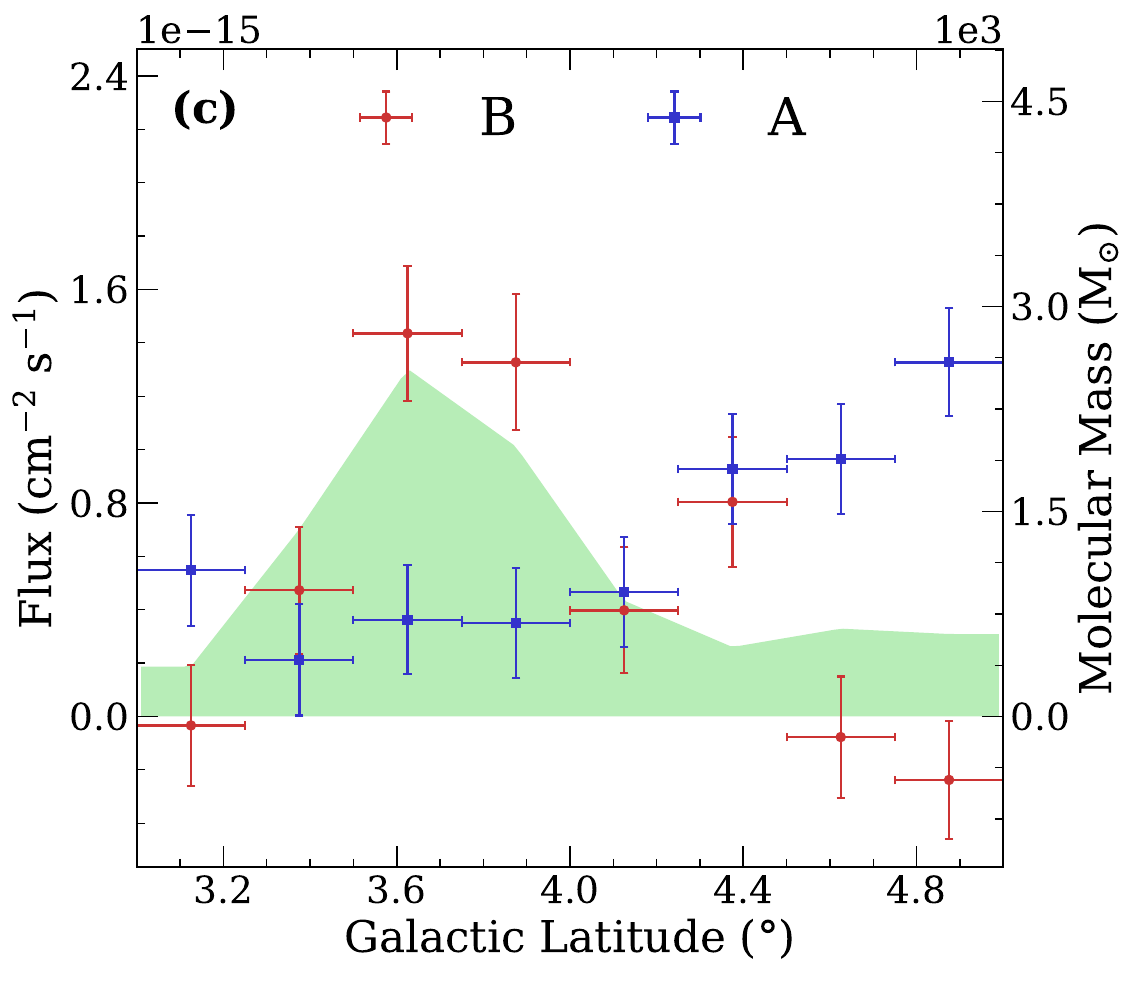}
%\hline
\includegraphics[width=0.32\textwidth]{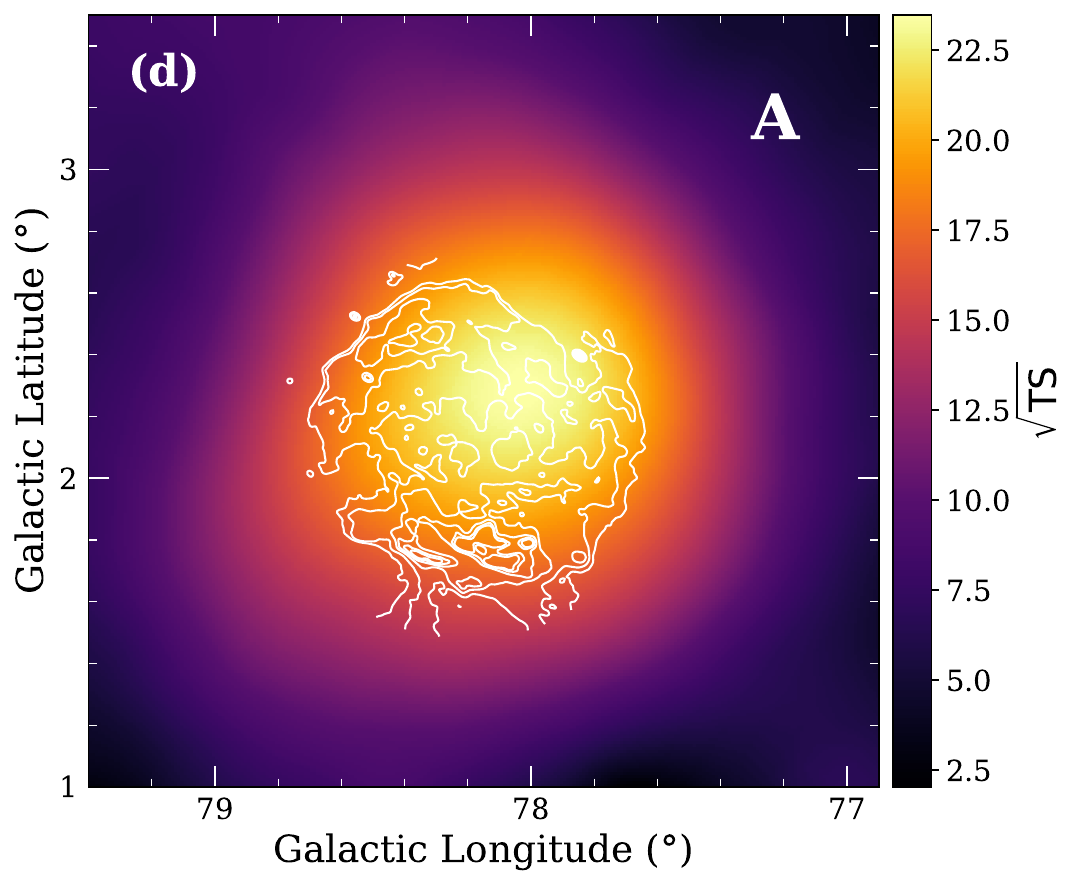}
\includegraphics[width=0.32\textwidth]{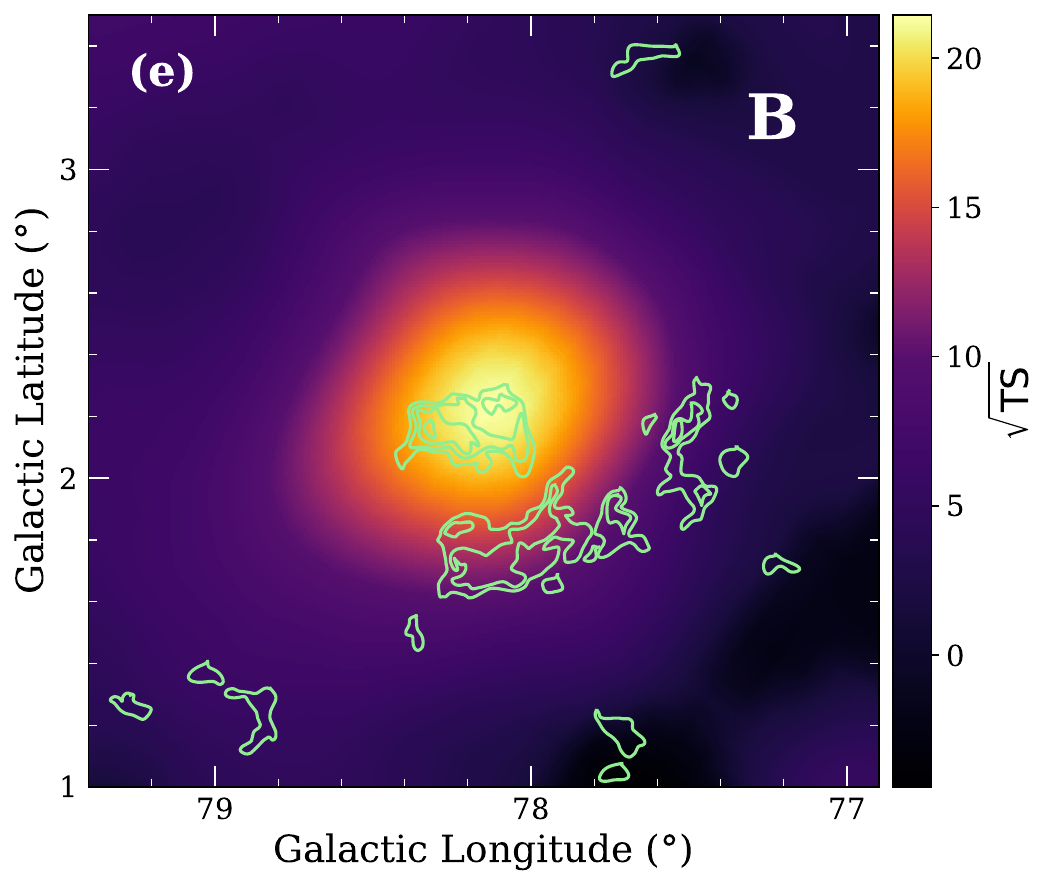}
\includegraphics[width=0.32\textwidth]{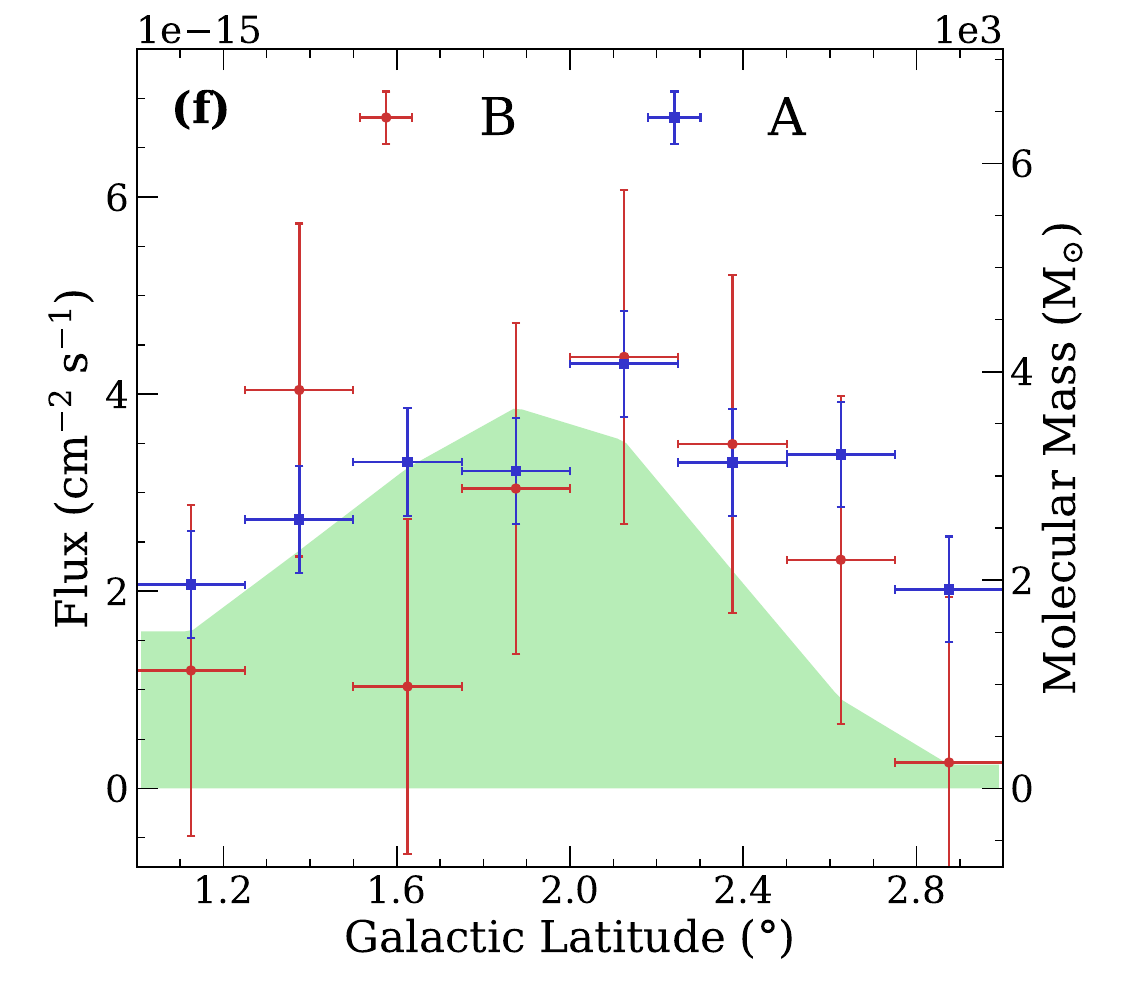}
\caption{\textbf{Significance maps of gamma-ray emission components for G150.3$+$4.5 (top) and $\gamma$-Cygni (bottom).}
Panels (a, b) and (d, e) correspond to components A and B, respectively. 
Radio continuum contours (white) \cite{2014A&A...566A..76G,2013MNRAS.436..968L} or CO molecular gas contours (green) \cite{2024A&A...686A.305F,2019ApJS..240....9S} are overlaid. 
CO emission is integrated over velocities of $-9$ to $-6$ km s$^{-1}$ for G150.3$+$4.5 \cite{2024A&A...686A.305F} and $-17$ to $-11$ km s$^{-1}$ for $\gamma$-Cygni \cite{2019ApJS..240....9S}. 
Panels (c, f) present 1D latitude profiles of integrated fluxes (blue: A; red: B) and molecular mass ($0.25^\circ$ belts) over longitudes $149.5^\circ$–$151.5^\circ$ (G150.3$+$4.5) and $77^\circ$–$79^\circ$ ($\gamma$-Cygni); fluxes of A are rescaled by $3\times10^{-3}$ and $10^{-2}$, respectively. 
}
\label{fig:sigmap_comp}
\end{figure}

Figure~\ref{fig:sigmap_comp} shows significance maps of different source components of
the two SNRs. 
Top panels are for SNR G150.3$+$4.5, and bottom panels are for
$\gamma$-Cygni. Radio, produced by high-energy electrons through the synchrotron process\cite{2014A&A...566A..76G,2013MNRAS.436..968L}, and molecular CO\cite{2024A&A...686A.305F,2019ApJS..240....9S} observations are over-plotted for 
comparison. It is evident that source A (low-energy component) of G150.3$+$4.5 is morphologically
consistent with the radio image of the SNR. The $39\%$ containment radius of the LHAASO 
emission is about $1.26^{\circ}$ (corresponding to a $68\%$ radius of approximately $1.9^{\circ}$), 
giving an emission size slightly larger than the radio extension ($2.5^{\circ}$ wide and 
$3.0^{\circ}$ high). Source B (high-energy component) shows a clear coincidence with the 
molecular cloud distribution at a similar distance of $\sim740$ pc with a total mass of $\sim 5\times 10^3M_\odot$\cite{2024A&A...686A.305F,2024ApJS..272...36F}, suggesting that the 
ultra-high-energy $\gamma$-ray emission may come from hadronic interactions of CRs with 
dense clouds. To better illustrate such a correlation, we divide the source 
region into a series of narrow slices of the Galactic latitude, with a latitude step size 
of $0.25^{\circ}$ and a longitude width of $2.0^{\circ}$, and derive the $\gamma$-ray fluxes 
of sources A and B, as well as the integrated CO mass in each sub-region. The results are 
shown in panel (c) of Figure~\ref{fig:sigmap_comp}. We can find a good correlation of the 
$\gamma$-ray fluxes with the CO traced masses for source B of SNR G150.3$+$4.5, supporting the hadronic 
origin of the $\gamma$-ray emission.

The case for $\gamma$-Cygni is similar. Source A (low-energy component) matches the radio and GeV $\gamma$-ray image of the SNR\cite{2023A&A...670A...8M}. 
Its extension is also slightly larger than that of the $0.5^{\circ}-0.6^{\circ}$ radio and GeV 
$\gamma$-ray disk\cite{2023A&A...670A...8M}. Source B (high-energy dominant) is significantly
smaller and is correlated with molecular clouds at a similar distance with a total mass of $\sim 10^4M_\odot$. The point-like source C 
is positionally consistent with sources detected by VERITAS\cite{2013ApJ...770...93A} 
and MAGIC\cite{2023A&A...670A...8M}, which have been interpreted as CR-illuminated 
molecular clouds\cite{2023A&A...670A...8M}. The results are shown in Figure \ref{fig:sigmap_pt} \cite{methods}. The $90\%$ upper limit of the extension of 
source C is $0.19^{\circ}$, which is consistent with those measured by VERITAS and 
MAGIC\cite{2013ApJ...770...93A,2023A&A...670A...8M}. 
Panel (f) shows the correlation of the LHAASO $\gamma$-ray fluxes and CO traced gas 
masses along Galactic latitudes from $1^{\circ}$ to $3^{\circ}$. To obtain this, we integrate over 
the longitude range from $77^\circ$ to $79^\circ$.

\begin{figure}[!htb]
\centering
\includegraphics[width=0.45\textwidth]{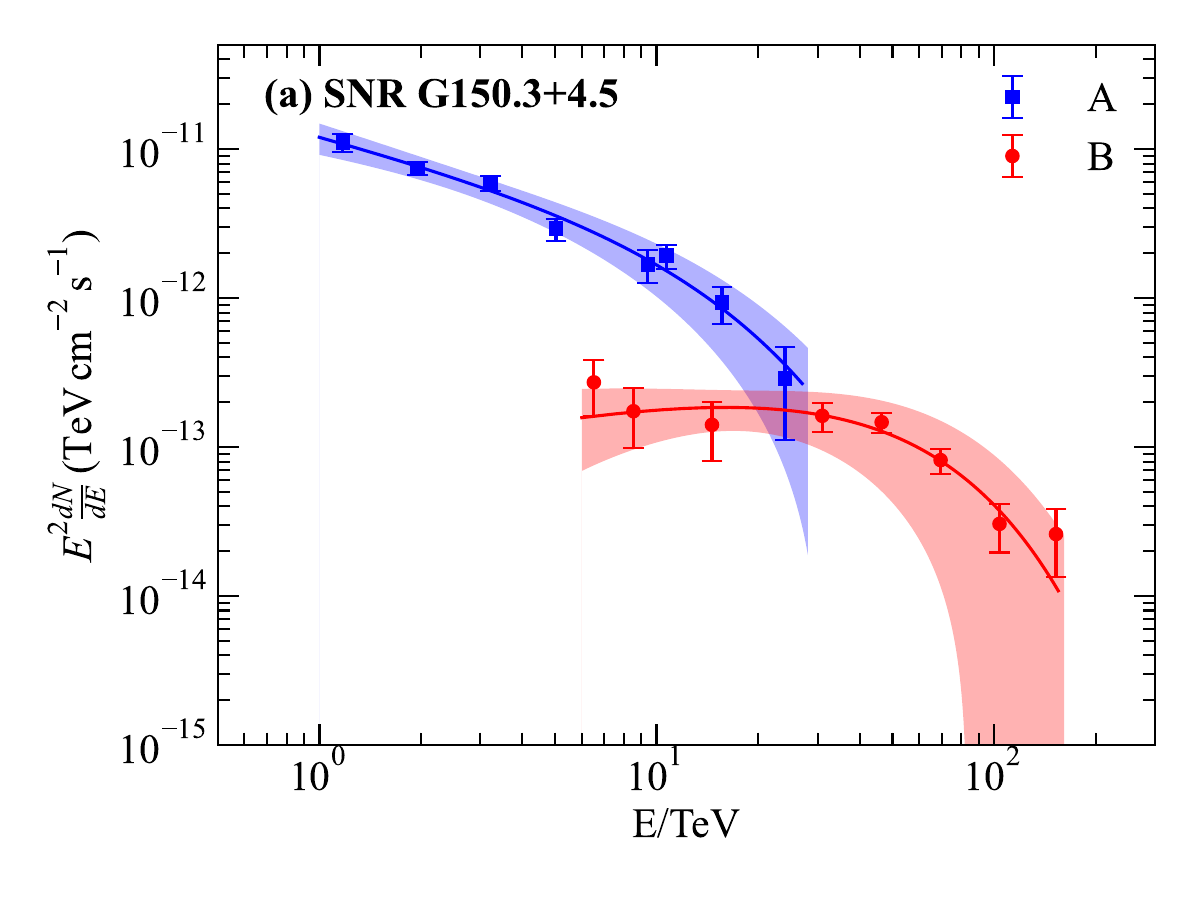}
\includegraphics[width=0.45\textwidth]{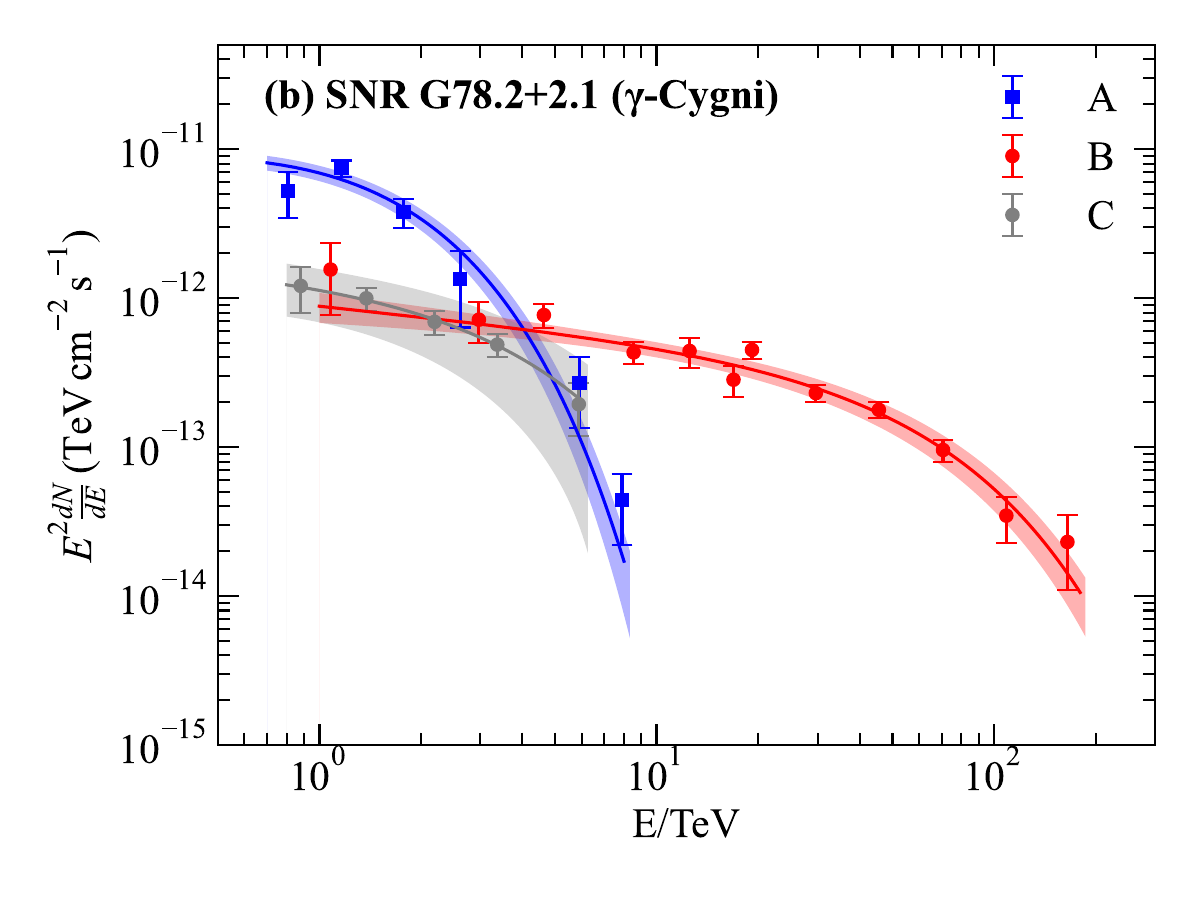}
\caption{\textbf{The SEDs of various components of SNR G150.3+4.5 (left) and $\gamma$-Cygni (right).}
The PLEcut fitting results and $68\%$ uncertainty ranges are shown by curves and hatched bands.}
\label{fig:SED}
\end{figure}

Figure~\ref{fig:SED} gives the spectral energy distributions (SEDs) of different source 
components of both SNRs. For each component the spectral shape is assumed to be a power-law 
with an exponential cutoff (PLEcut). We test the significance of the spectral cutoff, by 
comparing the fitting with a single power-law spectrum, and the increases in the TS value are 
given in Table \ref{tab:SED}. Except for source C of $\gamma$-Cygni, the spectral cutoff
is significant for all the other source components (the significance is approximately 
$\sqrt{\Delta{\rm TS}_{\rm cut}}$). The cutoff energies of source A are lower than those 
of source B for both SNRs. The correlation of source B with molecular 
clouds, together with the hard curved spectra, are consistent with the scenario that 
energetic particles accelerated early by SNR shocks escape from the acceleration sites 
and hit nearby clouds to produce ultra-high-energy $\gamma$ rays\cite{2007ApJ...665L.131G}. 
The highest energy photons from source B exceed 100 TeV, which indicates that the maximum 
energy of the accelerated protons reaches $\gtrsim$1 PeV, which offers evidence that the SNRs are CR 
PeVatrons in the Milky Way.

\begin{table}[htb]
\small
\centering
\caption{\textbf{
Fitting spectral parameters of various source components.} 
Columns from left are: source name, flux ($\phi_0$) at a reference energy of $5~\mathrm{TeV}$, 
spectral index ($\Gamma$), cutoff energy ($E_{\rm cut}$), and the increase of TS value compared 
with power-law fitting ($\Delta{\rm TS}_{\rm cut}$).}
\begin{tabular}{l|c|c|c|c} \hline \hline
Source name &$\phi_0$ ($\mathrm{TeV}^{-1}\mathrm{cm}^{-2}\mathrm{s}^{-1}$)  & $\Gamma$  & $E_{\rm cut}~(\mathrm{TeV})$ & $\Delta$TS$_{\rm cut}$ \\ \hline
SNR G150.3+4.5\textunderscore A & $(1.44\pm 0.33) \times 10^{-13}$  & $2.57 \pm 0.16$ & $13.38 \pm 5.32$ & 33\\  
SNR G150.3+4.5\textunderscore B & $(5.98 \pm 3.75 )\times 10^{-15}$ & $1.56 \pm 0.41$ & $36.20 \pm 12.01$ & 40 \\  
\hline
$\gamma$-Cygni\textunderscore A & $(1.05 \pm 0.37) \times 10^{-14} $ & $1.63 \pm 0.03$ & $1.03 \pm 0.07$ &60\\  
$\gamma$-Cygni\textunderscore B & $(2.31 \pm 0.30) \times 10^{-14}$ & $2.22 \pm 0.09$ & $54.40 \pm 5.15$ &41\\  
$\gamma$-Cygni\textunderscore C & $(1.12\pm0.87) \times 10^{-14} $ & $2.12 \pm 0.25$ & $3.34 \pm 1.21$ &22\\  
\hline 
\hline
\end{tabular}
\label{tab:SED}
\end{table}

\begin{figure}
\centering
\includegraphics[width=0.45\textwidth]{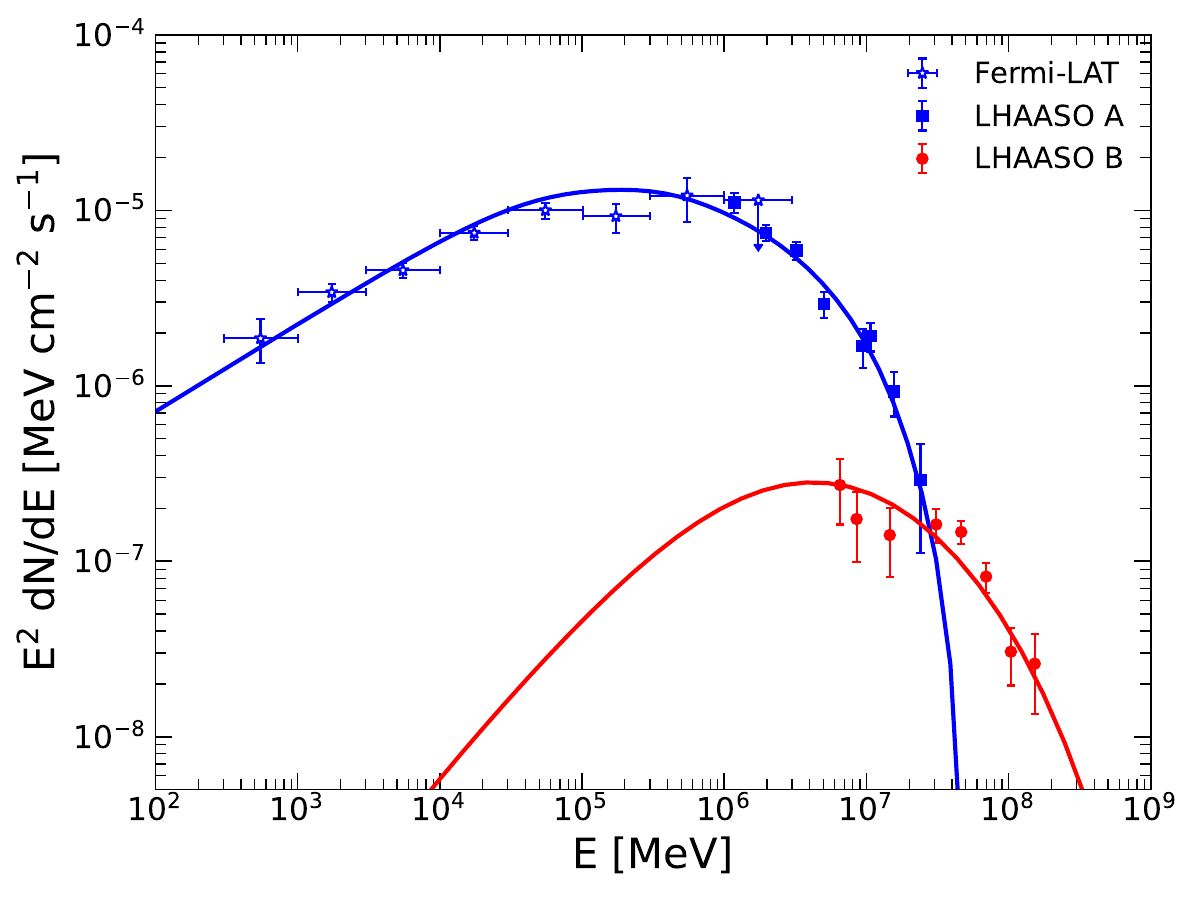}
\includegraphics[width=0.45\textwidth]{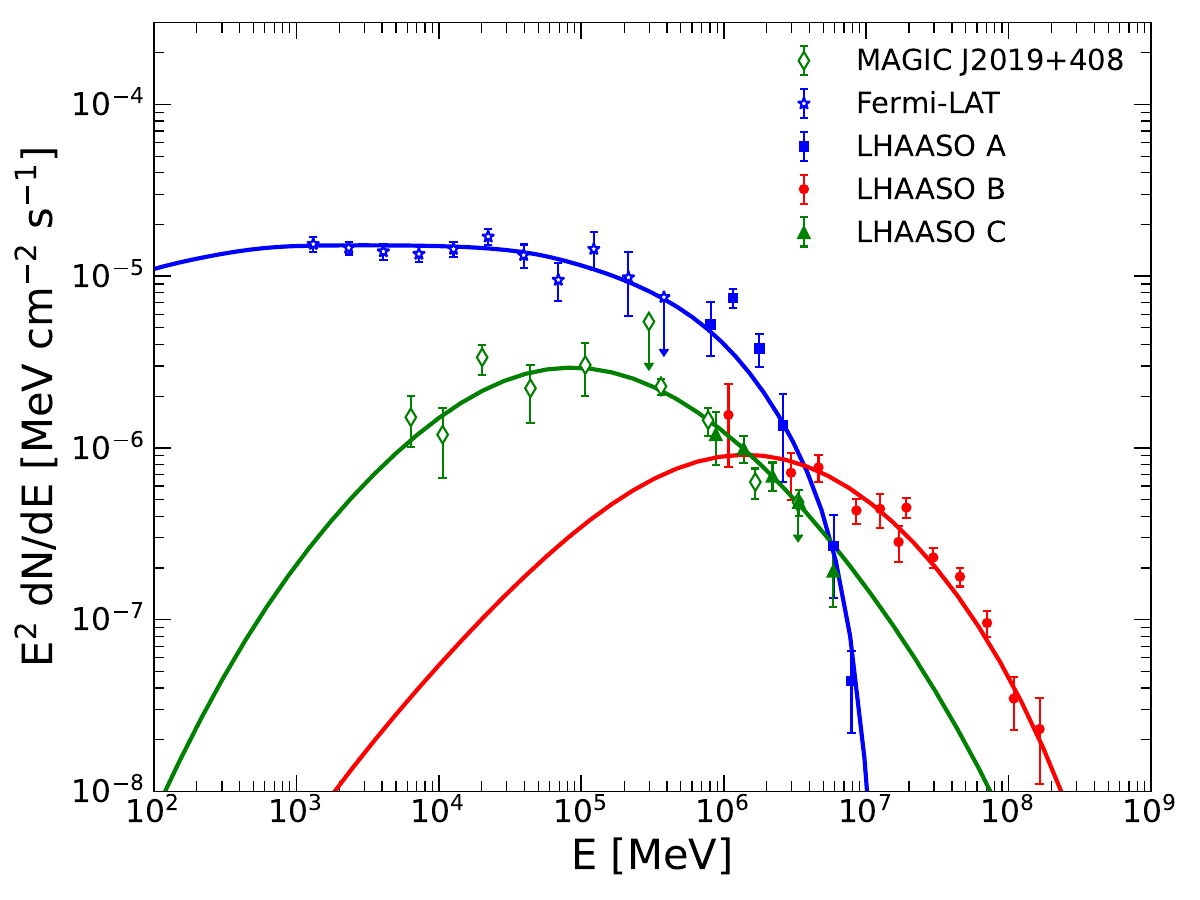}
\caption{{\bf Model fitting results of SEDs of G150.3+4.5 (left) and $\gamma$-Cygni (right).}
Fermi-LAT observations\cite{2020A&A...643A..28D,2018ApJ...861..134A} of these two SNRs at low 
energies and MAGIC observation of MAGIC J2019+408\cite{2023A&A...670A...8M} in the vicinity of
$\gamma$-Cygni are also shown. The blue line is a leptonic model to account for source A, and 
the red (green) line shows the $\pi^0$-decay contribution from escaping protons when hitting 
molecular clouds to account for source B (and C for $\gamma$-Cygni).}
\label{fig:model}
\end{figure}

\section*{Interpretation}

We consider a hybrid model containing both leptonic (inverse Compton scattering) and hadronic 
($\pi^0$ decay) $\gamma$-ray emission processes. 
We assume that high-energy CR protons and electrons were 
accelerated at an early stage of the SNR and were then injected instantaneously into the 
interstellar medium. These particles then diffuse out and interact with the ambient matter 
and fields, and produce multi-wavelength emission\cite{2011MNRAS.410.1577O,2012MNRAS.427...91O}.
A broken power-law with an exponential 
high-energy cutoff function\cite{2019ApJ...874...50Z,2019MNRAS.490.4317C} of the injection spectra for both 
electrons and protons is assumed. Efficient cooling of electrons results in a 
cutoff energy smaller than that of protons at present. The spatial and spectral distributions of electrons and 
protons are detailed in the Supplementary Materials \cite{methods}. 

The $\gamma$-ray emission of both sources A is explained by the inverse Compton scattering 
of injected electrons off the background radiation field. 
%cosmic microwave background (CMB) with a temperature ($T$) of $2.7$ K and an energy density ($n_{\gamma}$) of $0.26$ eV cm$^{-3}$, infrared (IR) background photon field with $T= 20$ K and $n_e= 0.3$ eV cm$^{-3}$ and optical photons field with $T= 5000$ K and $n_e= 0.3$ eV cm$^{-3}$, which have the significance of the spectral cutoff corresponding the electron cutoff energy $E^e_{cut}$ at 35 TeV and 9.1 TeV for G150.3+4.5 and $\gamma$-Cygni, respectively. 
%To get the total contribution of such a population of electrons, we integrate Eq. (\ref{eq:escape}) in the total volume of space from the present time to the birth of the SNR. The diffusion coefficient is assumed to be\cite{2017Sci...358..911A} $D_0=1.0 \times 10^{27}$ cm$^2$ s$^{-1}$ and $\delta=1/3$. 
We integrate the electrons in all the space to get the total spectrum, which is independent
of the diffusion coefficient. The spectral index $\alpha$ is fixed at 2.0 (3.0), as expected 
from the strong shock acceleration, below (above) a break energy of 3.0 TeV for G150.3+4.5 and 0.1 TeV for 
$\gamma$ -Cygni as determined by fitting to the data. Radiative cooling 
results in a cutoff energy smaller than the maximum injection energy, as shown by the solid 
lines in the Figure \ref{fig:elect_dist} \cite{methods}. %For G150.3+4.5 ($\gamma$-Cygni), the 
%cooling induced cutoff energy is found to be about 35 (9.1) TeV. 
The total energy of the electrons injected
 above $1$ GeV is found to be $W_e=2.3 \times 10^{47}$ erg for G150.3+4.5 and 
$W_e=1.6 \times 10^{49}$ erg for $\gamma$-Cygni. As shown by the blue lines in Figure 
\ref{fig:model}, the inverse Compton emission can very well account for Fermi-LAT and LHAASO 
source A from GeV to tens of TeV. The synchrotron radiation of the same electron population 
can also explain the radio emission, as detailed in the Supplementary Materials \cite{methods}(Figure
\ref{fig:multiwave}). Emission by electrons upstream of the shock, where the magnetic field 
is weak, may explain the slightly more extended $\gamma$-ray morphology than for the radio source.

The high-energy sources B (and also C for $\gamma$-Cygni) can be attributed to
high-energy CRs escaping from the SNR and illuminating ambient molecular clouds via inelastic
hadronic collisions with nuclei there. The normalization of the total injected proton
energy above $1$ GeV is fixed at $W_p=10^{50}$ erg. We find that a molecular cloud at a 
distance of $\sim$ 31 pc from the particle injection site and mass of $6.5 \times 10^3 
M_{\odot}$ can reproduce the LHAASO measurement of G150.3+4.5 well, for a slow diffusion 
coefficient $D_0=1.0 \times 10^{27}$ cm$^2$ s$^{-1}$ and $\delta=1/3$ as inferred by 
observations of pulsars\cite{2017Sci...358..911A}. For sources B and C of $\gamma$-Cygni, 
the corresponding distance and mass are $\sim15$ pc and $7.0 \times 10^4 M_{\odot}$, and 
$\sim 9.3$ pc and $3.5 \times 10^3M_{\odot}$, respectively. The cutoff energy of protons 
is estimated to be about 800 TeV for both SNRs, implying efficient PeV particle acceleration. 
The comparison of the model prediction and the data is shown in Figure \ref{fig:model}. 

SNR shocks have been considered to be the dominant agent for the acceleration of CR\cite{1976RvMP...48..161G}. 
The maximum acceleration energy of SNRs has been unclear for a long time. Although SNR W51 has been detected with LHAASO above 100 TeV\cite{2024SciBu..69.2833C}, it is compact with a strong X-ray pulsar wind nebula inside, implying leptonic contributions to TeV emission.
Ultra-high-energy $\gamma$ rays
produced by accelerated particles when hitting nearby molecular clouds, as illustrated by the detection of distinct high-energy components here by LHAASO, offer a direct probe of this problem. The highest energy of
$\gamma$-ray photons from dense cloud sites exceeds 100 TeV, strongly supporting the view that
SNR shocks can serve as PeVatrons to accelerate CR ions to PeV energies. However,
the spectral cutoff of the soft injection spectrum, characterized by an exponential cutoff energy
$E_{\rm cut}$, is found to be below PeV, indicating that SNR may not be dominant contributors to CR
well above the knee. The acceleration of CR by other sources, such as the center of the Milky
Way galaxy\cite{2016Natur.531..476H}, massive star clusters\cite{2019NatAs...3..561A}, and/or 
microquasars\cite{2024Natur.634..557A,2025NSRev..12af496L}, is possibly relevant to the
high-energy end of Galactic CR spectra.

\clearpage % Clear all remaining figures and tables then start a new page

% The list of references goes after the main text and before the acknowledgements
% When preparing an initial submission, we recommend you use BibTeX, like this:
%
\bibliography{science_template} % for a file named science_template.bib
\bibliographystyle{sciencemag}

% After the paper has completed peer review and been revised ready for acceptance,
% you should comment out the lines above and copy-paste the contents of your .bbl
% file here instead. This will help ensure that our conversion software works correctly.
% Remember to re-run BibTeX first - check the timestamp!
%
% Example of the first three entries copy-pasted from science_template.bbl:
%
%\begin{thebibliography}{1}
%
%\bibitem{example}
%A.~N. {Author}, An example reference. \emph{Journal of Improbable Research}
%  \textbf{1}, 67 (2020).
%
%\bibitem{example2}
%F.~M. {Surname}, S.~{Author}, A second example. \emph{Interesting Research
%  Letters} \textbf{32}, 897 (2019).
%
%\bibitem{example_preprint}
%P.~{One}, P.~{Two}, P.~{Three}, {An unpublished preprint}. \emph{preprint}
%  (2021), arXiv:2101.12345.
%
%\end{thebibliography}

%%%%%%%%%%%%%%%% ACKNOWLEDGEMENTS %%%%%%%%%%%%%%%

\section*{Acknowledgments}
We would like to thank all staff members who work at the LHAASO site above 4400 meter above
the sea level year round to maintain the detector and keep the water recycling system, electricity power supply and other components of the experiment operating smoothly. We are grateful to
Chengdu Management Committee of Tianfu New Area for the constant financial support for research
with LHAASO data. We appreciate the computing and data service support provided by the National High Energy Physics Data Center for the data analysis in this paper. This study made use of the data from the Milky Way Imaging Scroll Painting (MWISP) 
project, which is a multi- line survey in 12CO/13CO/C18O along the northern galactic 
plane with the PMO-13.7m telescope. We are grateful to all the members of the MWISP 
working group, particularly the staff members at the PMO-13.7m telescope, for their 
long-term support.
\paragraph*{Funding:}
This research work is
supported by the following grants: the National Natural Science Foundation 
of China (Nos. 12220101003,  12322302, 12573053, 12333006, 12405128), the Natural Science Foundation for General Program of 
Jiangsu Province of China (No. BK20242114), and the Jiangsu Provincial Excellent 
Postdoctoral Program (No. 2022ZB472), The National Natural Science Foundation of China No.12393851, No.12393852, No.12393853, No.12393854, 
No.12205314, No.12105301, No.12305120, No.12261160362, No.12105294, No.U1931201, No.12375107, NSFC No.12173039, the Department of Science and Technology of Sichuan
Province, China No.24NSFSC2319, Project for Young Scientists in Basic Research of Chinese Academy
of Sciences No.YSBR-061 and and in Thailand from the NSRF via the Research and Innovation Acceleration Agency for Competitiveness and Area Development (RCAD) (Program Management Unit for Technology and Innovation for Future Industries (PMU-B) : Brainpower for Future Industries) [grant number B39G690003]. MWISP was sponsored by National Key Research and Development Program 
of China with grants 2023YFA1608000, 2017YFA0402701, and by CAS Key Research Program 
of Frontier Sciences with grant QYZDJ-SSW-SLH047.
\paragraph*{Author contributions:}
%List each author’s contributions to the paper.
%Use initials to abbreviate author names.
Ying-Ying Guo analyzed the LHAASO data under the supervision 
of Yi Zhang, which were initiated by You-Liang Feng ($\gamma$-Cygni) and Han-Rong Wu 
(G150.3+4.5 WCDA) and cross-checked by Shi-Cong Hu. Hou-Dun Zeng, Qiang Yuan, Siming Liu, 
and Yi Zhang interpreted the results. Yang Su analyzed the CO data for $\gamma$-Cygni. 
Ye Li analyzed the ROSAT data for G150.3+4.5. Ying-Ying Guo, Qiang Yuan, Hou-Dun Zeng, 
and Siming Liu prepared the draft, with inputs and comments from other authors.\\
\textbf{Corresponding author emails: } yyguo@pmo.ac.cn; zhd@pmo.ac.cn; yuanq@pmo.ac.cn; 
liusm@swjtu.edu.cn; zhangyi@pmo.ac.cn; hushicong@ihep.ac.cn; wuhr@ihep.ac.cn and fengyouliang@utibet.edu.cn.

\paragraph*{Competing interests:}
%Disclose any potential conflicts of interest for all authors, such as patent applications,
%additional affiliations, consultancies, financial relationships etc.
%See the journal editorial policies web page for types of competing interest that must be declared.
%If there are no competing interests, state:
%``There are no competing interests to declare.''
The authors declare that they have no competing financial interests.

\paragraph*{Data and materials availability:}
%\textbf{Data Availability}
The authors declare that the data supporting the findings of this study are available within the paper, its supplementary information files, and the Large High Altitude Air-shower Observatory (https://english.ihep.cas.cn/lhaaso/pdl/202110/t20211026\_286779.html).
%\textbf{Code Availability} 
The findings presented in this study are derived exclusively from LHAASO  observations and standard LHAASO data analysis procedures. The analysis framework was developed by the LHAASO collaboration, and all codes used to obtain the results in this paper are publicly available on the website (https://english.ihep.cas.cn/lhaaso/pdl/202110/t20211026\_286779.html).
The simulation codes and data analysis codes are available from the corresponding 
authors upon reasonable request.
\subsection*{Supplementary materials}
%Materials and Methods\\
%Supplementary Text\\
%Figs. S1 to S3\\
%Tables S1 to S4\\
LHAASO experiment.\\
LHAASO data analysis.\\
Multi-wavelength observations.\\
MWISP CO observations.\\
Spectral modeling.\\
%Supplementary Text\\
Figures S1 to S4\\
Tables S1\\
References \textit{(39-\arabic{enumiv})}\\ % automatically fills out the last reference number
% (filling out the other numbers automatically is possible but fiddly and liable to break)
%Movie S1\\
%Data S1

%%%%%%%%%%%%%%%% END OF MAIN TEXT %%%%%%%%%%%%%%%

\newpage

%%%%%%%%%%%%%%%% START OF SUPPLEMENT %%%%%%%%%%%%%%%

% Figures, tables, equations and pages in the supplement are numbered S1, S2 etc.
\renewcommand{\thefigure}{S\arabic{figure}}
\renewcommand{\thetable}{S\arabic{table}}
\renewcommand{\theequation}{S\arabic{equation}}
\renewcommand{\thepage}{S\arabic{page}}
\setcounter{figure}{0}
\setcounter{table}{0}
\setcounter{equation}{0}
\setcounter{page}{1} % not 0 as \newpage already started a supplementary page
% References continue the numbering from the main text.

%%%%%%%%%%%%%%%% SUPPLEMENT TITLE PAGE %%%%%%%%%%%%%%%

\begin{center}
\section*{Supplementary Materials for\\ \scititle}

% Author list for the supplement
% Indicate the corresponding authors, but do NOT include institutions here
% It would be nice if the template auto-generated this, but doing so is complicated...
%First~Author$^{\ast\dagger}$,
%A.~Scientist$^\dagger$,
%Someone~E.~Else\\ % we're not in a \author{} environment this time, so use \\ for a new line
%\small$^\ast$Corresponding author. Email: example@mail.com\\
%\small$^\dagger$These authors contributed equally to this work.
\end{center}

% Fill out the numbers for each type of supplementary material,
% and delete any lines that aren't applicable.
% These are just example numbers that don't match the rest of this template.
\subsubsection*{This PDF file includes:}
LHAASO experiment.\\
LHAASO data analysis.\\
Multi-wavelength observations.\\
MWISP CO observations.\\
Spectral modeling.\\
%Supplementary Text\\
Figures S1 to S4\\
Tables S1\\
%Captions for Movies S1 to S2\\
%Captions for Data S1 to S2

%\subsubsection*{Other Supplementary Materials for this manuscript:}
%Movies S1 to S2\\
%Data S1 to S2

\newpage

%%%%%%%%%%%%%%%% MATERIALS AND METHODS %%%%%%%%%%%%%%%

%\subsection*{Methods}

\subsection*{LHAASO experiment.}
LHAASO is a hybrid, large area, wide field-of-view observatory for CRs and $\gamma$ rays in a 
wide energy range. It is located at Haizi Mountain ($100^{\circ}\!.01$E, $29^{\circ}$\!.35N; 
4400 m above the sea level), Daocheng, Sichuan province, China\cite{2022ChPhC..46c0001M}.
LHAASO consists of three detector arrays, the square kilometer array (KM2A), the water 
Cherenkov detector array (WCDA), and the wide field-of-view Cherenkov telescope array (WFCTA). 
The KM2A consists of 5195 electromagnetic detectors (EDs) placed on a 15 m spacing grid and 
1188 muon detectors (MDs) placed on a 30 m spacing grid. An outer ring of EDs with 30 m spacing 
is built to separate the in-array and out-array showers. The total area of KM2A is about 
1.3 km$^2$. The KM2A gives the most sensitive survey of the northern sky in ultra-high-energy 
$\gamma$ rays\cite{2021ChPhC..45b5002A}. 
The WCDA has 3120 water Cherenkov detectors placed evenly in three pools with a depth of 4.4 m
and a total area of 78,000 m$^2$. Each detector unit has a dimension of $5\times5$ m$^2$, 
covered with non-reflecting black plastic curtains. Two photo-multipliers are equipped in each
unit to record the Cherenkov light produced by charged particles in the pure water. The WCDA is
sensitive to $\gamma$ rays from 100 GeV to 30 TeV\cite{2021ChPhC..45h5002A}.
The WFCTA consists of 18 telescopes, each with an area of 4.7 m$^2$ and a field-of-view of
$16^{\circ}\times16^{\circ}$. The telescopes are placed in the center of the KM2A, which are
used to image the showers to precisely measure the composition of CRs.

\subsection*{LHAASO data analysis.}
In this analysis, the WCDA and KM2A data about the two SNRs are used. The WCDA data used 
are the full array data from March, 2021 to July, 2024, with a live time of 1136.2 days.
The zenith angle of the reconstructed direction is chosen to be less than 50$^{\circ}$.
We adopt the PINCness parameter (Parameter for Identifying Nuclear Cosmic-rays), defined 
as\cite{2017ApJ...843...39A}
\begin{equation}
P = \frac{1}{N}\sum_{i=0}^{N} \frac{({\zeta}_i- \langle {\zeta}_i \rangle)^2}{{\sigma_{\zeta_i}}^2},
\end{equation}
to select photon-like events and reject the CR background. Here $\zeta_i=\log_{10} (Q_i)$, 
$Q_i$ is the hit charge, $\sigma_{\zeta_i}$ is the width of the distribution of $\zeta_i$ 
obtained from simulated $\gamma$-ray data. In this work PINCnes is required to be less than 
%1.12, 1.02, 0.90, 0.88, 0.84 and 0.84 
1.02, 0.90, 0.88, 0.88, 0.84 and 0.84
for different energy bands with number of hits values
of [60–100), [100–200), [200–300), [300–500), [500–800), and [800–2000].

The KM2A data used were collected by the 1/2 array with a live time of about 289.6 days, the 
3/4 array with a live time of about 215.9 days, and the full array with a live time of about 
1064.9 days. The muon-to-electromagnetic-particle ratio, $R_{\mu e}=\log[(N_{\mu}+10^{-4})/N_e]$, 
is employed to suppress the CR background. The cuts of the $R_{\mu e}$ parameter is optimized
according to the observation of the Crab nebula\cite{2021ChPhC..45b5002A}. Further selections 
of the KM2A data include:
\begin{itemize}
\item the number of triggered EDs and the number of deposited particles used in the shower
reconstruction are both larger than 10; 
\item the zenith angle of the reconstructed direction is less than 50$^{\circ}$; 
\item the number of particles detected within 40 m from the shower core is larger than that 
within $40-100$ m; 
\item the shower age is within $0.6-2.4$.
\end{itemize}

We use the direct integral method\cite{2004ApJ...603..355F} to estimate the background.
This method assumes that the spatial distribution of the detection efficiency is stable
in a reasonably short time period. The background is then determined by the observed
events in adjacent time bins. A time step of 4 hours is adopted, and at each step a time 
window of $\pm5$ hours is used to estimate the background. 

The 3D-likelihood method, a joint analysis of the morphology and spectrum, is employed to 
simultaneously fit the data from the WCDA and KM2A detectors. The sky around the target source 
is binned into cells with a size of $0^{\circ}\!.1$ in both the right ascension (R.A.) and 
declination (Dec.). A likelihood-ratio test is performed, with the test statistic (TS) 
defined as TS~$=2\ln({\mathcal L}_{s+b}/{\mathcal L}_b)$, where ${\mathcal L}_{s+b}$ is 
the maximum likelihood for the signal plus background hypothesis and ${\mathcal L}_b$ is 
the likelihood for the background only hypothesis. In the background only case, the TS value 
obeys a $\chi^2$ distribution with $n$ degrees of freedom, where $n$ is the number of additional
parameters in the signal plus background model. The statistical significance of the detection 
of a source can be obtained from the Wilks’s theorem\cite{1938wilks}. 

Since both sources are located in the Galactic plane, it is necessary to properly take 
into account the Galactic diffuse emission (GDE). In the analysis of G150.3+4.5, we 
simultaneously fit the target source and the GDE within a $4^{\circ}$ circle region 
centered at (R.A. $=67^{\circ}$, Dec. $=55^{\circ}$), where the spectrum of the GDE is 
modeled as a broken power-law function as ref.\cite{2025PhRvL.134h1002C} and the morphology 
is modelled with the Planck dust opacity template\cite{2016A&A...596A.109P}. For the 
analysis of $\gamma$-Cygni, the region of interest (ROI) is a circle with a radius of 
$5^{\circ}$ centered at (R.A. $=307^{\circ}$, Dec. $=41^{\circ}$), with the bright source 
MGRO J2019+37 (VER J2019+368) masked out by a disk of $2.5^{\circ}$
radius\cite{2007ApJ...658L..33A,2018ApJ...861..134A}. 
%There are three main sources in the ROI near $\gamma$-Cygni, TeV J2032+4130 which is 
%a PWN\cite{2005A&A...431..197A}, the Cygnus
%Cocoon\cite{2011Sci...334.1103A,2014ApJ...790..152B,2021NatAs...5..465A}, and the 
%$\gamma$-Cygni SNR\cite{2013ApJ...770...93A,2023A&A...670A...8M,2018ApJ...861..134A}. 
%These sources overlap with each other, and can account for most of the TeV emission 
%observed in the ROI.
The $\gamma$-Cygni region predominantly overlaps with the Cygnus bubble, which is modeled 
using a Gaussian template for the inner part (i.e., the Cocoon region) and gas templates 
derived from the HI4PI 21-cm line survey\cite{2016A&A...594A.116H} and the CfA galactic 
CO survey\cite{2001ApJ...547..792D}, consistent with the methodologies described in
ref.\cite{2011Sci...334.1103A}. Since the flux of the Cygnus bubble within the ROI is 
significantly higher than the GDE flux\cite{2024SciBu..69..449L}, we find that the 
contamination from the GDE to the analysis of $\gamma$-Cygni is negligible.

\subsection*{Multi-wavelength observations.}

G150.3$+$4.5 is a shell-type SNR. The eastern part of the shell was initially discovered 
by Gerbrandt et al. (2014)\cite{2014A&A...566A..76G}, showing faint, extended non-thermal 
radio emission, which was named as G150.8$+$3.8. Gao and Han (2014)\cite{2014A&A...567A..59G} 
reported the total shell-like morphology with a size of $2.5^\circ \times 3.0^\circ$ 
using the $\lambda$ 6 cm survey data. The non-thermal radio spectral indices for the 
eastern and western shells of G150.3$+$4.5 are $\sim -0.4$ and $\sim -0.7$, 
respectively\cite{2014A&A...567A..59G}. The hard power-law $\gamma$-ray emission of 
G150.3$+$4.5 was discovered by \cite{2016PhDT.......190C} through analyzing seven years 
of Pass 8 data recorded by the Fermi-LAT. The $\gamma$-ray morphology was found to be 
extended with a radius of about $1.4^\circ$. Then in Fermi-LAT catalogs 
(2FHL\cite{2016ApJS..222....5A}, FGES\cite{2017ApJ...843..139A}, 4FGL\cite{2022ApJS..260...53A}),
an extended source with about $1.5^\circ$ radius was reported together with a hard spectrum 
in GeV band with a power-law spectral index of about 1.9. Devin et al. (2020)\cite{2020A&A...643A..28D} 
used a two-dimensional symmetric Gaussian model or a disk model, spatially coincident with 
the radio emission, to describe the $\gamma$-ray emission using more than 10 years of 
Fermi-LAT data, and obtained a hard spectral index of $\Gamma=1.62$ at $E_0=9.0$ GeV 
with a spectral curvature toward high energies.
An unidentified source with a steep spectrum in the east-southern part of the shell,
4FGL J0426.5+5434, was present in the 4FGL catalog\cite{2022ApJS..260...53A}. This 
source is a point-like source positionally coincident with LHAASO source B. No pulsation 
is reported for this source. We regard it as a background source, due to its distinct 
spectrum and morphology from the LHAASO source.
The ROSAT X-ray observation of this region gave only an upper limit of the non-thermal
component assuming a power-law spectrum with $\Gamma=2$\cite{2020A&A...643A..28D}. 
We re-analyze the data and derive an unabsorbed non-thermal $0.1-3.0$ keV flux upper 
limit (95\%) of about $9.0 \times 10^{-11}$ erg cm$^{-2}$ s$^{-1}$ for a power-law 
index of $\Gamma=3.0$ which is closer to the model prediction and a maximum HI column 
density of $3.97\times10^{21}$ cm$^{-2}$. 
Feng et al. (2024)\cite{2024A&A...686A.305F} presented large-field CO line observations 
toward SNR G150.3+4.5 using the 13.7 m millimeter telescope of Purple Mountain Observatory. 
Via the measurement of the molecular clouds associated with the SNR, the distance of the
SNR was estimated to be about 740 pc, and the age was $(1-18) \times 10^4$ year. Another 
estimate of the age according to the statistical diameter-age relation\cite{2023ApJS..265...53R}
gave an age of $\sim26$ kyr. In this work, we adopt 20 kyr as the age of G150.3$+$4.5 for 
the modeling.

SNR $\gamma$-Cygni is also a shell-type SNR with a radio diameter of
$\sim1^{\circ}$\cite{2013MNRAS.436..968L}. Gao et al. (2011)\cite{2011A&A...529A.159G} 
observed this SNR in radio bands at five wavelengths, and gave a power-law spectral index 
of $-0.48\pm0.03$. The ROSAT observations revealed shell structures compatible 
with the radio emission, and showed a thermal spectrum\cite{2013MNRAS.436..968L}. 
No evidence of non-thermal radiation of the SNR was found in X-ray band except weak 
emission from the nebula of PSR J2021+4026 in the center of the SNR was observed by 
Chandra and XMM-Newton\cite{2015ApJ...799...76H}. 
%$3.6 \times 10^6$ photons kev$^{-1}$ cm$^{-2}$ s$^{-1}$ at 1 keV with a photon index of $\Gamma =1.5 \pm 0.8$. %$8.6 \times 10^{-14}$ erg cm$^{-2}$ s$^{-1}$ in $0.5-10$ keV.
%Uchiyama et al. (2002)\cite{2002ApJ...571..866U} performed observations of this SNR with \textit{ASCA} and found two clumps of hard X-ray emission in the north-west, of which one is likely of extra-galactic origin and other one may be a hard X-ray point source\cite{2013MNRAS.436..968L}. 
Using the ASCA data, Aliu et al. (2013)\cite{2013ApJ...770...93A} derived an upper limit of 
the $0.5-8.0$ keV unabsorbed flux from the region of VER J2019+407 of $1.9\times10^{-12}$ 
erg cm$^{-2}$ s$^{-1}$, for an assumed power-law component with $\Gamma=2.0$. 
In $\gamma$-ray band, $\gamma$-Cygni has been detected by Fermi-LAT, VERITAS, MAGIC, and 
HAWC\cite{2013ApJ...770...93A,2018ApJ...861..134A,2023A&A...670A...8M,2020ApJ...905...76A}. 
The Fermi-LAT observation showed extended emission likely associated with the radio shell, 
and a less extended emission in the north-western part possibly associated with the TeV source
MAGIC J2019+408\cite{2023A&A...670A...8M} which is possibly conterpart of the VERITAS source
VER J2019+407\cite{2013ApJ...770...93A,2018ApJ...861..134A}. Emission associated with the radio 
shell was also detected by MAGIC\cite{2023A&A...670A...8M}. In the third source catalog of HAWC, 
3HWC J2020+403 was detected and possibly associated with VER J2019+407\cite{2020ApJ...905...76A}.

\subsection*{MWISP CO observations.}
\begin{figure}[!htp]
\includegraphics[width=0.33\linewidth]{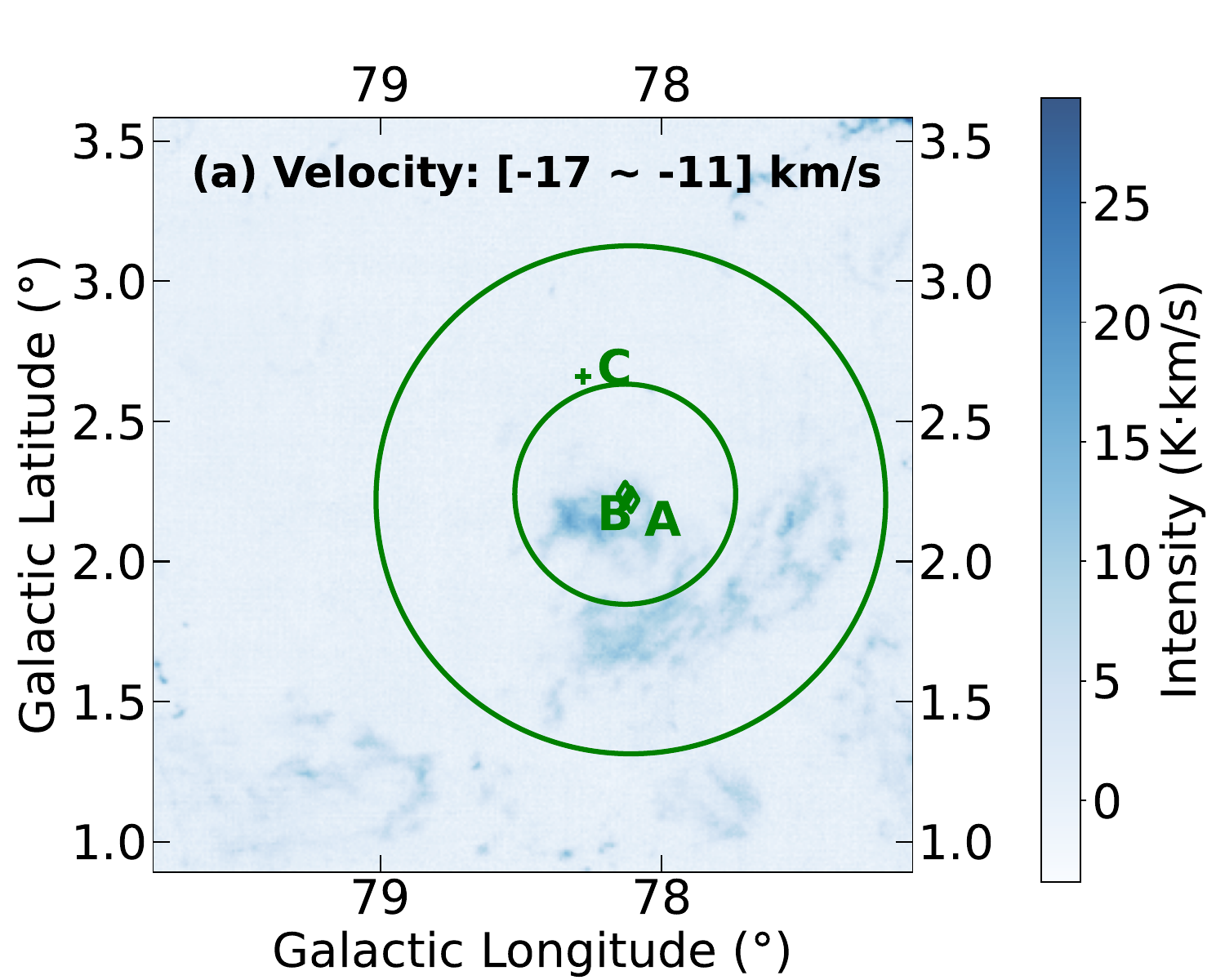}
\includegraphics[width=0.33\linewidth]{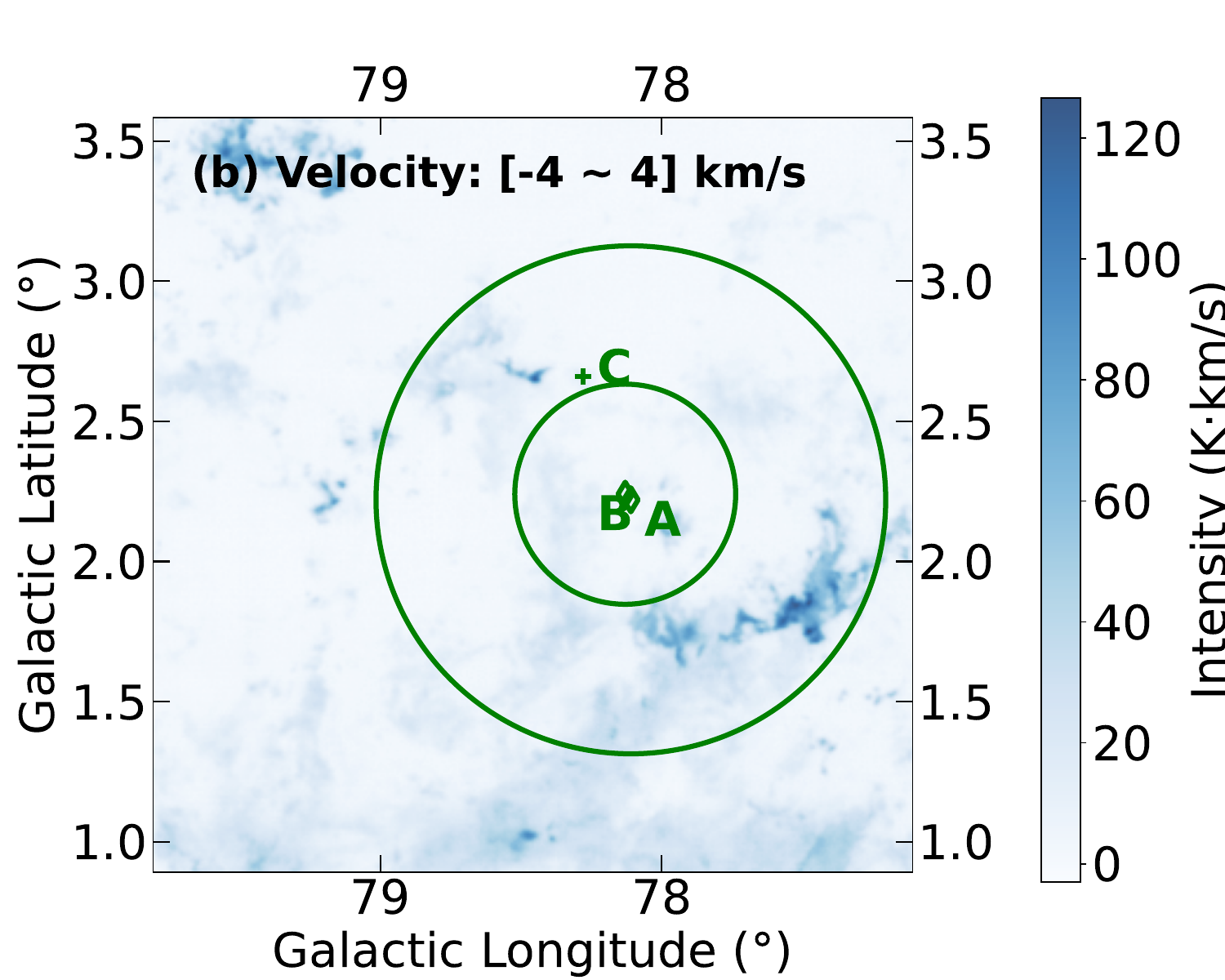}
\includegraphics[width=0.33\linewidth]{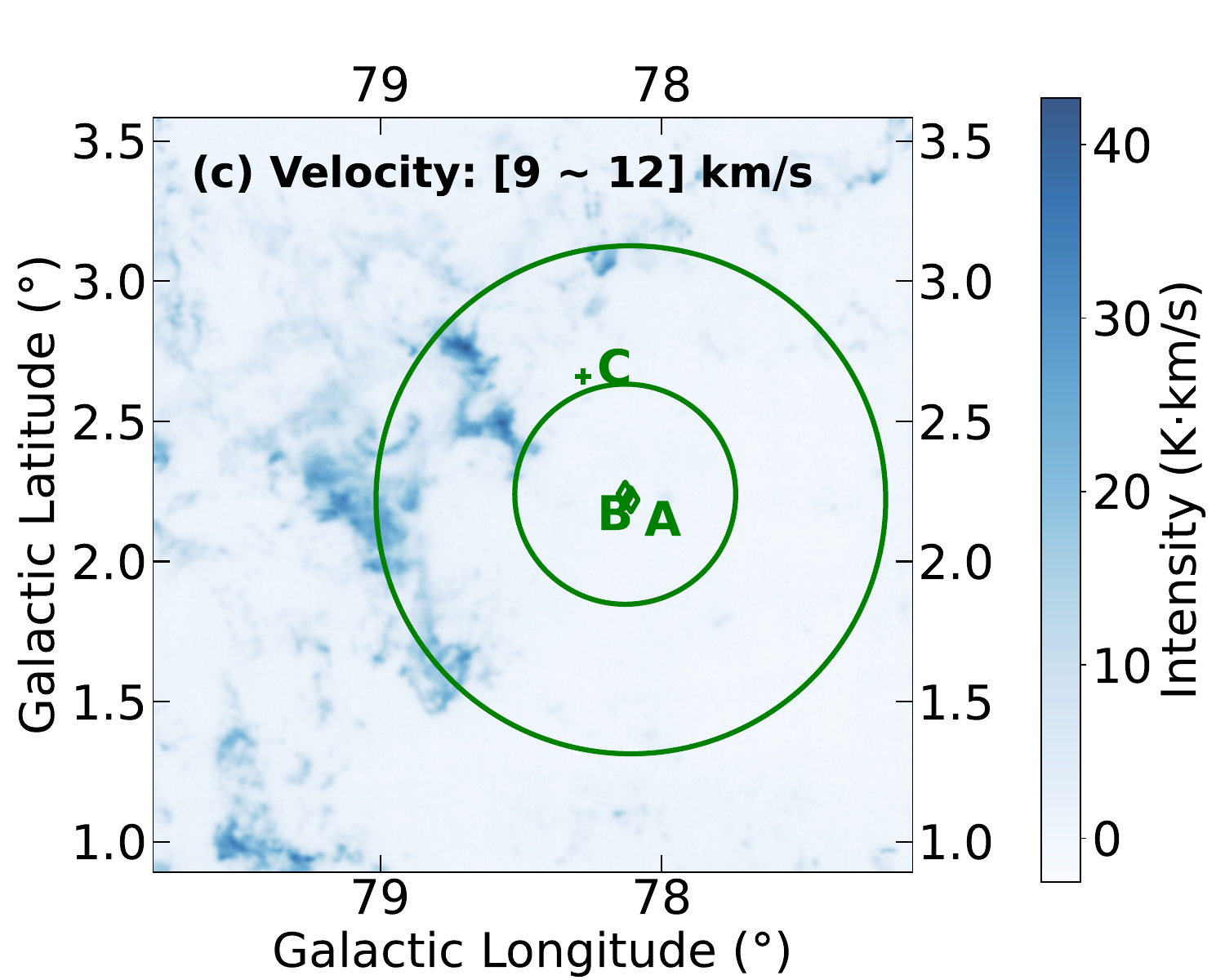}
\caption{{\bf Integrated CO emission (color maps) towards $\gamma$-Cygni in different velocity ranges.}
The positions of LHAASO sources A, B, and C are marked in green, along with their $68\%$ extension 
regions.
}
\label{fig:G78CO}
\end{figure}

\begin{figure}[!htb]
\centering
\includegraphics[width=0.32\textwidth]{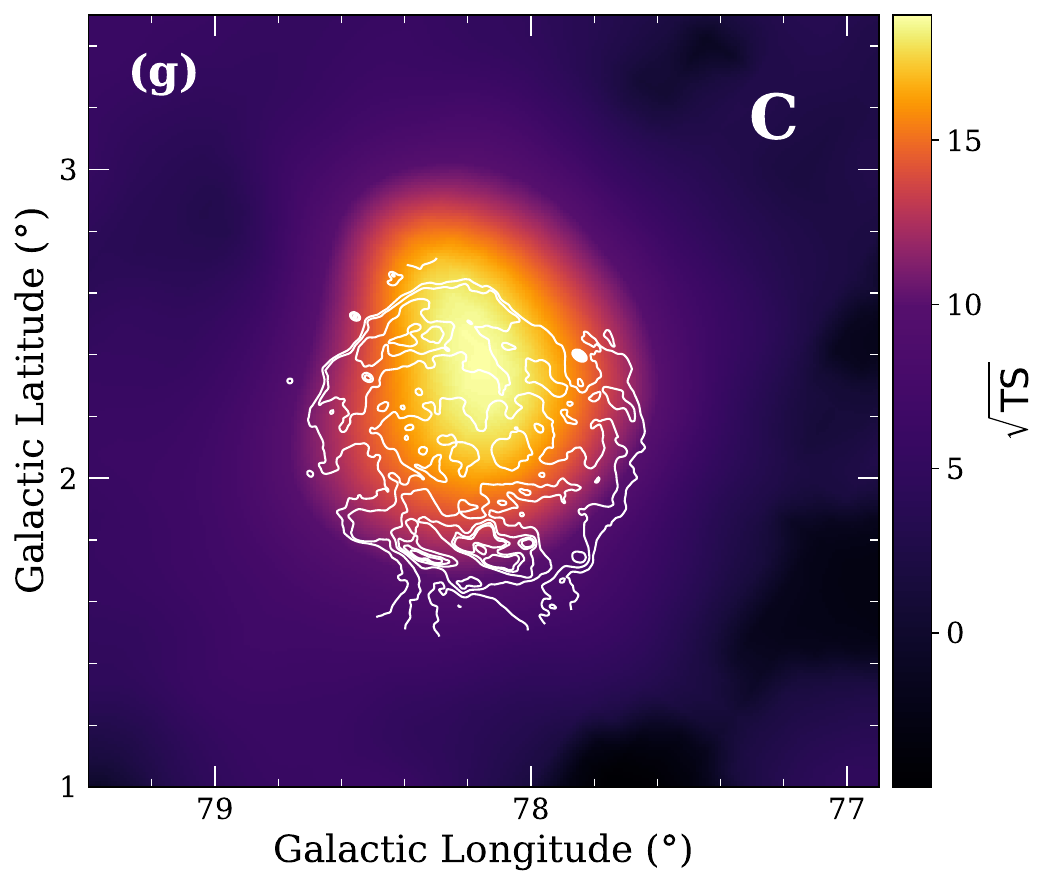}
\includegraphics[width=0.32\textwidth]{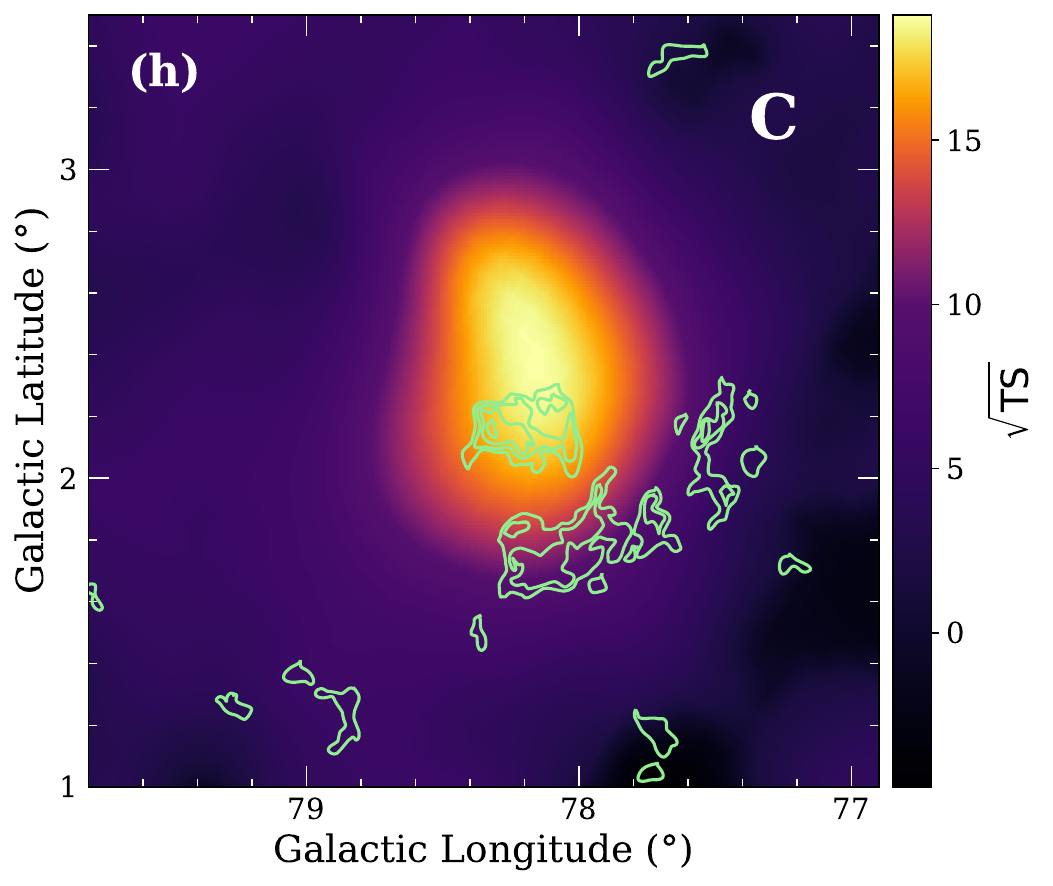}
\caption{\textbf{Significance maps of gamma-ray emission components for the component C of $\gamma$-Cygni.}
Radio continuum contours (white) \cite{2013MNRAS.436..968L} and CO molecular gas contours (green) \cite{2019ApJS..240....9S} are overlaid. 
CO emission is integrated over velocities of $-17$ to $-11$ km s$^{-1}$. 
}
\label{fig:sigmap_pt}
\end{figure}

Information about the molecular cloud distributions in sky regions around the two SNRs comes 
from the CO observations by the Milky Way Imaging Scroll Painting (MWISP) 
project\cite{2019ApJS..240....9S}. For G150.3$+$4.5, we direct use the results from a recent 
publication of Feng et al. (2024)\cite{2024A&A...686A.305F}, which report a distance to the SNR of approximately 740 pc and a mass for the MC G150.6+3.7 of about $5 \times 10^3$ solar masses. This cloud is spatially coincident with source B of LHAASO's results.
For $\gamma$-Cygni, we analyze 
the CO emission data observed by the MWISP. 
%Briefly, the $^{12}$CO($J$=1--0), $^{13}$CO($J$=1--0), and C$^{18}$O ($J$=1--0) lines are simultaneously observed by using the 13.7 m millimeter-wavelength telescope at Qinghai observation station. The half-power beam-width of the telescope is $\sim50''$ and the three-dimensional (3D) FITS data cubes were made with a grid spacing of $30''$. Typically, the sensitivity of the data is $\sim$0.5~K for $^{12}$CO and $\sim$0.3~K for $^{13}$CO and C$^{18}$O at a velocity resolution of about 0.2~km~s$^{-1}$. 
The SNR $\gamma$-Cygni is located in the Cygnus X region and there are a large number of OB stars. The CO gas toward the region is concentrated in three velocity ranges: $<-10$~km~s$^{-1}$, $-9$ to $+6$~km~s$^{-1}$, and $>+7$~km~s$^{-1}$. The total molecular mass in a projected area of $\sim 3$ deg$^2$ is less than $2 \times 10^5$ solar masses. We note that the CO gas in the velocity ranges of $-17$ to $-11$~km~s$^{-1}$ (a), 
$-4$ to $+4$~km~s$^{-1}$ (b), and $+9$ to $+12$ ~km~s$^{-1}$ (c) is probably related to the 
SNR, as shown in Figure \ref{fig:G78CO}. The CO emission of the (a) component 
is relatively weak but displays a partial-shell structure in the southwest. The intense CO 
emission of the local arm (b) and the tangent point (c) traced by the other two components 
are widely distributed around the remnant. No prominent CO broadenings are found, indicating 
that the interactions between the SNR and molecular clouds should be weak. However, we find 
shell-like structures (or partial-shell structures) of molecular gas surrounding the SNR. 
The SNR $\gamma$-Cygni seems to evolve in a cavity-like structure with a relatively low 
density and the remnant just starts to interact with the surrounding dense clouds. 
This scenario can explain the absence of strong X-ray emission in the SNR's interior. 
From the spatial correlation, we expect that the molecular gas component (a) may be relevant 
to the LHAASO $\gamma$-ray source. 

\iffalse
The red color is for velocity from $+$9 to $+$12~km~s$^{-1}$, green for $-4$ to $+4$~km~s$^{-1}$, 
and blue for $-$17 to $-$11~km~s$^{-1}$. Also shown are massive OB stars with spectral type
earlier than B1 at distances of $1.5-2.0$~kpc (purple squares), the Effelsberg 11~cm radio 
continuum emission\cite{1990A&AS...85..691F} (white contours; start from 3400~mK and increase 
by a step of 1400~mK), the TeV source VER J2019+407\cite{2013ApJ...770...93A} (yellow circle), 
the Fermi-LAT source associated with the radio shell\cite{2023A&A...670A...8M} (cyan circle), 
PSR J2021$+$4026 (cyan dot), and LHAASO sources A, B, and C. 
{\bf (Please update LHAASO sources, as Figure 1.)}
\fi

\subsection*{Spectral modeling.}
We consider instantaneous injection of particles from a point-like source. The resulting 
distribution of particles is\cite{1995PhRvD..52.3265A}
\begin{equation}
   N_i(E_i,R,T)= \frac{\dot{P}(E_{i,0})Q_i(E_{i,0})}{\dot{P}(E_i)\pi^{3/2}R_{{\rm diff},i}^3}~\exp\left[-\frac{R^2}{R_{{\rm diff},i}^2}\right], 
   \label{eq:escape}
\end{equation}
where $i$ represents electron or proton, $E_{i,0}$ is the initial energy of a particle at 
injection time that cools down to $E_i$ at time $T$, $\dot{P}(E)$ is the energy loss rate, $R$ is the 
distance from the injection location (assumed to be the center of the SNR), and $T$ is 
the time from the injection (assumed to be approximately the age of the SNR). For protons 
the energy loss can be neglected, and we have $E_i \equiv E_{i,0}$. The diffusion length is
\begin{equation}
   R_{{\rm diff},i}=2\times\left\{
   \begin{array}{ll}
   \left[\int_{E_i}^{E_{i,0}}{D(E')}{\dot{P}^{-1}(E')}dE'\right]^{1/2}, & i={\rm electron,} \\
   \left[D(E_i)T\right]^{1/2}, & i={\rm proton.}
   \end{array}\right.
   \label{eq:rdif}
\end{equation}
where $D(E)=D_0 (\frac{E}{100~\rm{TeV}})^{\delta}$ is the diffusion coefficient.
The injection spectrum is assumed to be a double power-law function with an expotential 
high-energy cutoff\cite{2019ApJ...874...50Z} 
%\begin{equation}
%    N_i(E_i)=N_0^i E_i^{-\alpha}\left(1+E_i/E_{\rm br}\right)^{-1} \exp
%    \left(-E_i/E_{\rm cut}\right),
%\end{equation}
\begin{eqnarray}
\label{eq:Np}
Q_i(E_i)= Q_0^i~\textrm{exp}\left(-E_i/E_{\rm cut}\right)\times
\left\{
\begin{array}{lcl}
E_i^{-\alpha}&
& { \textrm{if}~~~ E_i < E_{\rm br} } \\
 E_{\rm br} E_i^{-(\alpha+1)}  &
& { \textrm{if}~~~E_i \geq E_{\rm br} }\;,
\end{array}
\right.
\end{eqnarray}
where $Q_0^i$ is the normalization, $\alpha$ is the low-energy spectral index, $E_{\rm br}$ 
is the break energy, and $E_{\rm cut}$ is the cutoff energy. 
%Note that, we assume the spectrum is steeper by one power at high energies, which is consistent with the prediction of ion-neutral collisions and the resulting Alfven wave evanescence\cite{2011NatCo...2..194M}. 

To calculate the cooling rate of electrons, the synchrotron cooling with a mean magnetic field
$B$ and the inverse Compton scattering (ICS) cooling in the radiation field of three gray-body
components\cite{2017Sci...358..911A}: the cosmic microwave background with temperature of
2.725 K and energy density of $0.26$ eV cm$^{-3}$, an infrared background with temperature of
20 K and energy density of $0.3$ eV cm$^{-3}$, and an optical background with temperature of
5000 K and energy density of $0.3$ eV cm$^{-3}$. The ICS cooling rate is calculated using the
parameterization\cite{2021ChPhL..38c9801F} based on the Klein-Nishina cross section.

\begin{figure}
\centering
\includegraphics[width=0.7\textwidth]{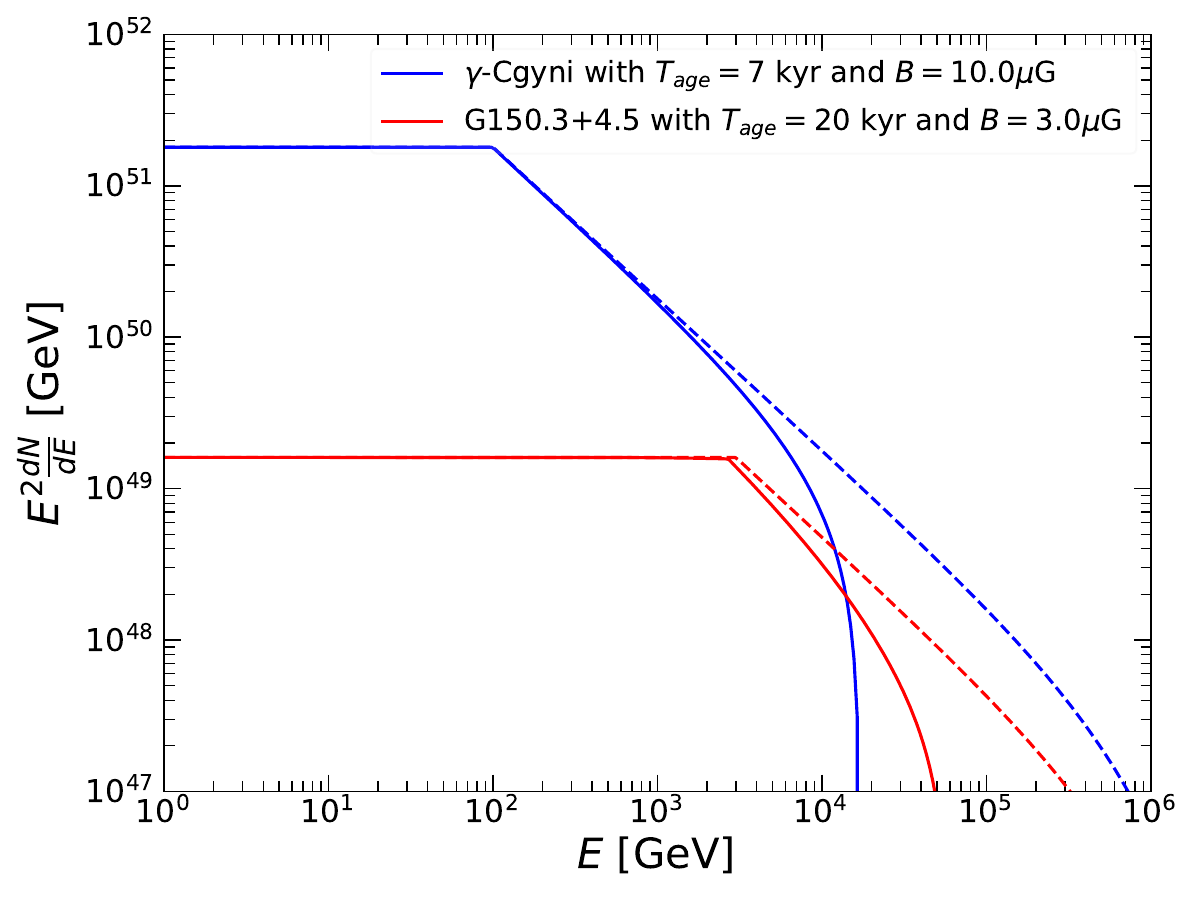}
\caption{{\bf The electron spectral distributions of G150.3+4.5 (red) and $\gamma$-Cygni (blue).} 
Dashed lines represent the injection spectra, characterized by a double power-law distribution 
with an exponential cutoff at 800 TeV. Deviations of solid lines from dashed ones show the
effect of radiative cooling of electrons.}
\label{fig:elect_dist}
\end{figure}

\begin{figure}
\centering
\includegraphics[width=0.45\textwidth]{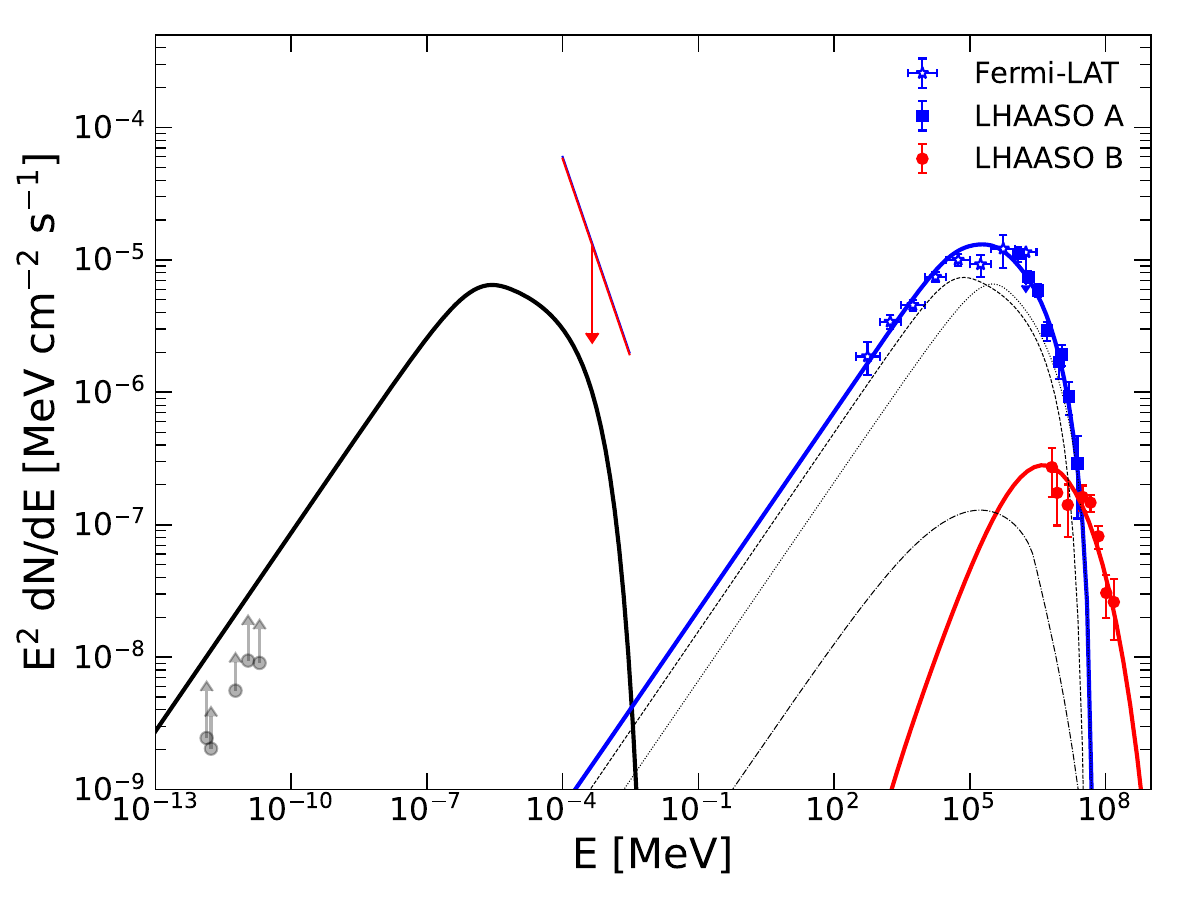}
\includegraphics[width=0.45\textwidth]{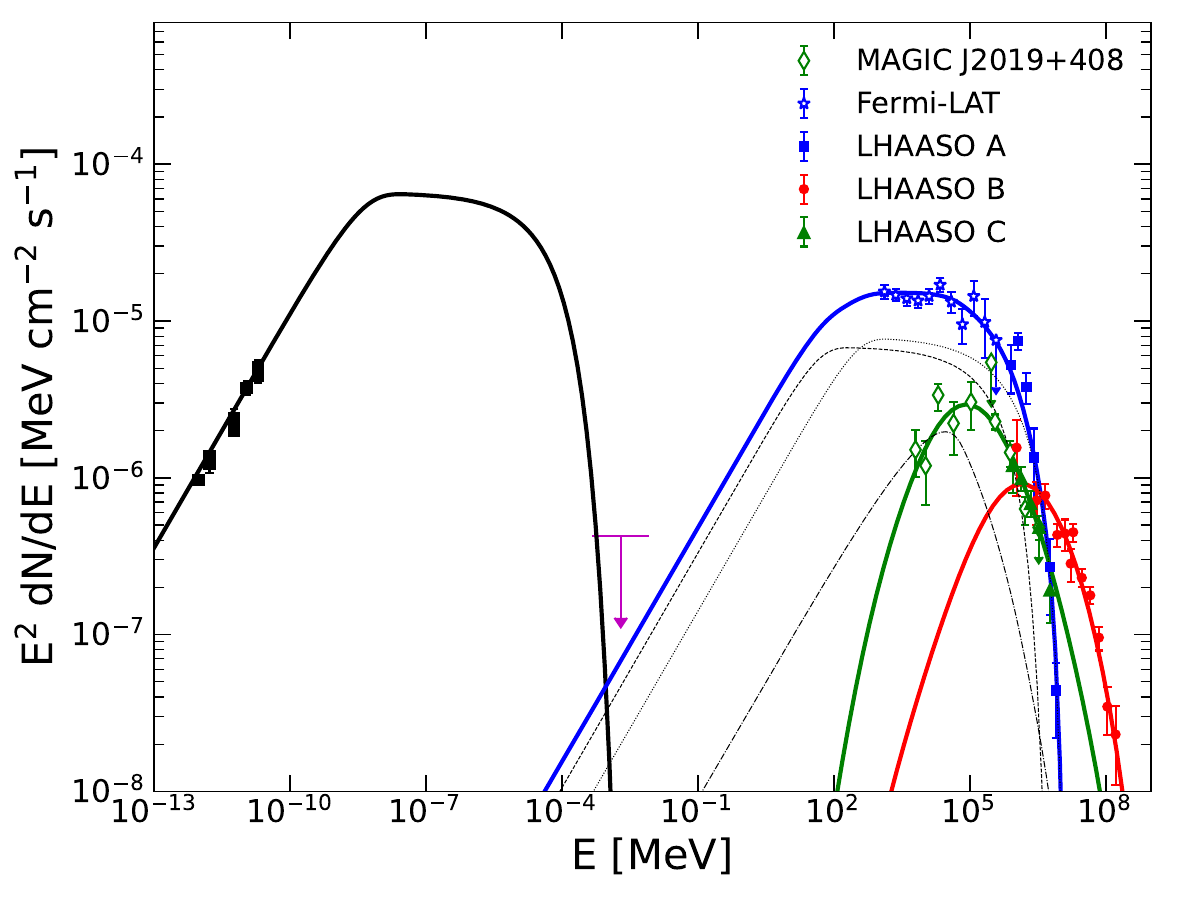}
\caption{{\bf Multi-wavelength SEDs of G150.3+4.5 (left) and $\gamma$-Cygni (right).} 
In each panel, the black solid line is the model predicted synchrotron radiation, and the
blue solid line is the total ICS emission. The ICS emission for different background
radiation components are shown, respectively, by black dashed (microwave), dotted (infrared), 
and dash-dotted (optical) lines. Red and green lines correspond to the hadronic emission
from escaping protons hitting molecular clouds. 
Measurements in the radio\cite{2014A&A...566A..76G,2011A&A...529A.159G}, X-ray\cite{2013ApJ...770...93A}, 
and $\gamma$-ray\cite{2020A&A...643A..28D,2018ApJ...861..134A,2023A&A...670A...8M,2021NatAs...5..465A} 
bands are employed to compare with the model prediction.
%with a mass of $6.5 \times 10^3 \times \frac{10^{50}}{W_p}$ M$_{\odot}$ at approximately $31$ pc from the acceleration site. The radio fluxes, which are the lower limits, come from Gerbrandt et al.\cite{2014AA...566A..76G} and the Fermi-LAT data is from Devin et al.\cite{2020A&A...643A..28D}. And this upper limit of X-rays is what we got by reanalyzing the ROSAT data. Right: the  red and green lines represent the presence of two molecular clouds with masses of $7.0 \times 10^4 \times \frac{10^{50}}{W_p}$ and $3.5 \times 10^3 \times \frac{10^{50}}{W_p}$ M$_{\odot}$, respectively, approximately $15$ pc and $9.3$ pc from the acceleration site. The radio and Fermi-LAT data are from Gao et al.\cite{2011A&A...529A.159G} and Abeysekara et al.\cite{2018ApJ...861..134A}. The gamma-ray of MAGIC J2019+408 from MAGIC Collaboration et al.\cite{2023A&A...670A...8M} and the X-ray uplimit from Aliu et al.\cite{2013ApJ...770...93A}. The HAWC measurement is indicated with the gray-shaded butterfly (Abeysekara et al.\cite{2021NatAs...5..465A}).
}
\label{fig:multiwave}
\end{figure}

The larger sources A of both SNRs are attributed to the ICS of electrons. The total electron
spectrum is calculated through an integration of Eq. (\ref{eq:escape}) in space. Via fitting 
to the multi-wavelength data, the injection source parameters obtained are $\alpha=2.0$, 
$E_{\rm br}=3.0$ TeV for G150.3+4.5 and 0.1 TeV for $\gamma$-Cgyni. The magnetic field 
strengths are about 3 $\mu$G for G150.3+4.5 and 10 $\mu$G for $\gamma$-Cgyni. The resulting 
electron spectra are shown by solid lines in Figure \ref{fig:elect_dist}. 
Compared with the injection spectra (dashed lines), the cooling effect results in an earlier 
cutoff of the electron spectra. For the parameters adopted in this work, the cutoff energies
are about 30 TeV for G150.3+4.5 and 10 TeV for $\gamma$-Cgyni. The multi-wavelength data-model 
comparison is shown in Figure \ref{fig:multiwave}.

Sources B (and also C for $\gamma$-Cygni) are interpreted as $\pi^0$ decay emission when 
escaping protons hit molecular clouds at specific distances. The $\gamma$-ray yield from 
proton-proton inelastic collisions is adopted from Ref.\cite{2014PhRvD..90l3014K}.
The injection spectra of protons shares the same form as those of electrons (dashed lines
in Figure \ref{fig:elect_dist}), but with different energy budgets. 
To account for the high-energy spectra measured by LHAASO, the injection spectra require
a cutoff at $\sim800$ TeV. The low-energy fluxes of protons would be suppressed due to the
in-efficient propagation, which result in relatively narrow spectra of $\gamma$-ray emission, 
as shown in Figure~\ref{fig:model}. All parameters used in the SED fitting for both SNRs are summarized in Table \ref{tab:fitted parameters}.

\begin{table} % Do not use \begin{table*}
    \scriptsize
	\centering
	% Captions go above tables
	\caption{\textbf{The injection spectral parameters and other parameters used in fitting for G150.3+4.5 and $\gamma$-Cygni.}}
	\label{tab:fitted parameters} % give each table a logical label name
	\begin{tabular}{l|ccccccc|c} % four columns, alignment for each
%		\\
		\hline
        \hline
		Source name &$\alpha$ & $E_{br}$[TeV] &$E_{cut}$[PeV] & $W_e$[erg] & $B [\mu$ G] & Age[kyr] & $D$[kpc]& Molecular clouds [M$_{\odot}$] \\
		\hline
		G150.3+4.5 & 2.0 & 3.0 &  0.8& $2.3 \times 10^{47}$ & $3.0$ & 20 & 0.74 &$5.0 \times 10^3 \frac{W_p}{10^{50} \rm{erg}}$ for B with R = 31 pc \\
        \hline
		\multirow{2}{*}{$\gamma$-Cygni}  & \multirow{2}{*}{2.0} & \multirow{2}{*}{$0.1$} &  \multirow{2}{*}{0.8}&\multirow{2}{*}{$1.6 \times 10^{49}$}& \multirow{2}{*}{$10.0$}& \multirow{2}{*}{7.0} & \multirow{2}{*}{1.7} &$7.0 \times 10^4 \frac{W_p}{10^{50} \rm{erg}} $\rm{\,for\, B \, with\, R = 15\, pc}\\
		&&&&&&&&$3.5 \times 10^3 \frac{W_p}{10^{50} \rm{erg}} $\rm {\,for \,C \,with\, R=9.3 \,pc} \\
		\hline
	\end{tabular}
\end{table}

Leptonic origin of the $\gamma$-ray emission may not be firmly ruled out. For $\gamma$-Cygni, 
sources A and B may be attributed to a halo powered by PSR J2021+4026, which has a relatively 
high spin-down luminosity of $\sim 10^{35}$ erg s$^{-1}$, although its PWN emission is 
weak and its overall $\gamma$-ray spectrum is distinct from other known pulsar halos\cite{2015ApJ...799...76H}. Source C can also be explained with the ICS emission of 
energetic electrons.
Source B of G150.3+4.5 may be attributed to a hidden pulsar as implied by a puzzling 
unidentified soft-spectrum $\gamma$-ray point source 4FGL J0426.5+5434 which is not detected 
at other wavelengths. 

Source B in G150.3+4.5 may also be associated with molecular clouds inside the radio shell of the SNR. In this case, the emission is produced by CRs that is still trapped in the SNR and may have already experienced significant energy loss due to adiabatic expansion in the Sedov phase of SNR evolution. The maximum energy therefore can go beyond a few PeV.

\clearpage
Zhen Cao$^{1,2,3}$,
F. Aharonian$^{3,4,5,6}$,
Y.X. Bai$^{1,3}$,
Y.W. Bao$^{7}$,
D. Bastieri$^{8}$,
X.J. Bi$^{1,2,3}$,
Y.J. Bi$^{1,3}$,
W. Bian$^{7}$,
J. Blunier$^{9}$,
A.V. Bukevich$^{10}$
% ...
\\
$^{1}$ State Key Laboratory of Particle Astrophysics \& Experimental Physics Division \& Computing Center, Institute of High Energy Physics, Chinese Academy of Sciences, 100049 Beijing, China\\
$^{2}$ University of Chinese Academy of Sciences, 100049 Beijing, China\\
$^{3}$ TIANFU Cosmic Ray Research Center, 610000 Chengdu, Sichuan,  China\\
$^{4}$ University of Science and Technology of China, 230026 Hefei, Anhui, China\\
$^{5}$ Yerevan State University, 1 Alek Manukyan Street, Yerevan 0025, Armeni a\\
$^{6}$ Max-Planck-Institut for Nuclear Physics, P.O. Box 103980, 69029  Heidelberg, Germany\\
$^{7}$ Tsung-Dao Lee Institute \& School of Physics and Astronomy, Shanghai Jiao Tong University, 200240 Shanghai, China\\
$^{8}$ Center for Astrophysics, Guangzhou University, 510006 Guangzhou, Guangdong, China\\
$^{9}$ APC, Universit\'e Paris Cit\'e, CNRS/IN2P3, CEA/IRFU, Observatoire de Paris, 119 75205 Paris, France\\
$^{10}$ Institute for Nuclear Research of Russian Academy of Sciences, 117312 Moscow, Russia\\
% ...

\end{document}